\documentclass[manuscript,screen,acmsmall,nonacm]{acmart}

\usepackage{xargs}      
\usepackage{soul}       
\usepackage{xcolor}     
\usepackage{enumitem}   
\usepackage{subcaption} 
\usepackage{multirow}   
\usepackage{algorithm}
\usepackage{algorithmic}
\usepackage{graphicx}

\usepackage{listings}
\lstdefinelanguage{json}{
    basicstyle=\tiny\ttfamily,
    stepnumber=1,
    numbersep=8pt,
    showstringspaces=false,
    breaklines=true,
    frame=none,
    literate=
     *{0}{{{\color{blue}0}}}{1}
      {1}{{{\color{blue}1}}}{1}
      {2}{{{\color{blue}2}}}{1}
      {3}{{{\color{blue}3}}}{1}
      {4}{{{\color{blue}4}}}{1}
      {5}{{{\color{blue}5}}}{1}
      {6}{{{\color{blue}6}}}{1}
      {7}{{{\color{blue}7}}}{1}
      {8}{{{\color{blue}8}}}{1}
      {9}{{{\color{blue}9}}}{1}
      {:}{{{\color{black}:}}}{1}
      {,}{{{\color{black},}}}{1}
      {\{}{{{\color{orange}\{}}}{1}
      {\}}{{{\color{orange}\}}}}{1}
      {[}{{{\color{orange}[}}}{1}
      {]}{{{\color{orange}]}}}{1},
}
\definecolor{darkgreen}{rgb}{0.0, 0.5, 0.0}  
\lstdefinelanguage{markdown}{
    basicstyle=\tiny\ttfamily,
    morekeywords={\#, \*\*, \*, -}, 
    keywordstyle=\color{blue},
    sensitive=false,
    morecomment=[l]{>}, 
    commentstyle=\color{gray}\itshape,
    morestring=[b]"
}

\newcommand{\crowdgenui}{\textsc{CrowdGenUI}}

\AtBeginDocument{%
  }

\begin{document}



\title{\crowdgenui{}: Aligning LLM-Based UI Generation with Crowdsourced User Preferences}


\author{Yimeng Liu}
\authornote{Work done during an internship at Adobe Research}
\affiliation{%
  \institution{University of California, Santa Barbara}
  \city{Santa Barbara}
  \country{USA}}
\email{yimengliu@ucsb.edu}

\author{Misha Sra}
\affiliation{%
  \institution{University of California, Santa Barbara}
  \city{Santa Barbara}
  \country{USA}}
\email{sra@ucsb.edu}

\author{Chang Xiao}
\affiliation{%
  \institution{Adobe Research}
  \city{San Jose}
  \country{USA}}
\email{cxiao@adobe.com}

\renewcommand{\shortauthors}{Liu et al.}


\begin{abstract}
Large Language Models (LLMs) have demonstrated remarkable potential across various design domains, including user interface (UI) generation. However, current LLMs for UI generation tend to offer generic solutions that lack a nuanced understanding of task context and user preferences. We present \crowdgenui{}, a framework that enhances LLM-based UI generation with crowdsourced user preferences. This framework addresses the limitations by guiding LLM reasoning with real user preferences, enabling the generation of UI widgets that reflect user needs and task-specific requirements. We evaluate our framework in the image editing domain by collecting a library of 720 user preferences from 50 participants, covering preferences such as predictability, efficiency, and explorability of various UI widgets. A user study (N=78) demonstrates that UIs generated with our preference-guided framework can better match user intentions compared to those generated by LLMs alone, highlighting the effectiveness of our proposed framework. We further discuss the study findings and present insights for future research on LLM-based user-centered UI generation.
\end{abstract}





\maketitle


\section{Introduction}
\label{sec:introduction}

Graphical user interfaces (GUIs) have long served as the primary medium through which users interact with software applications and digital systems~\cite{photoshop, illustrator}. Despite their ubiquity and effectiveness, designing intuitive and efficient GUIs remains a complex and labor-intensive process. It often requires expert knowledge, iterative user testing, and significant time investments---both from designers to refine the UI and from users to learn how to use it effectively.

In response, there has been growing interest in natural language interfaces, which offer an alternative mode of interaction by allowing users to express their intentions through language. These interfaces reduce the design and learning overhead, simplifying how users interact with complex tools~\cite{dalle, descript, maddigan2023chat2vis}. However, while convenient, natural language interfaces often fall short in scenarios requiring precision and control, such as adjusting image brightness or fine-tuning the parameters of a filter. Moreover, they typically lack the interactive, real-time feedback that GUIs provide, which is essential for tasks that benefit from trial-and-error and visual confirmation.

With recent advances in generative AI, particularly the emergence of large language models (LLMs), researchers have started exploring how LLMs might bridge this gap by automatically generating adaptive user interfaces~\cite{vaithilingam2024dynavis, cheng2024biscuit}. These systems aim to combine the high-level task understanding and domain flexibility of LLMs with the precise, user-driven control of traditional GUIs. Yet, current LLM-based UI generation methods rely mainly on the model’s internal reasoning and general knowledge. As a result, the generated UIs often appear generic and may fail to capture the nuanced requirements of specific tasks or the varied preferences of individual users.

To address these limitations, we present \crowdgenui{}, a novel framework that enhances LLM-based UI generation by incorporating crowdsourced user preferences. This framework leverages collective insights from real users to guide the LLM in generating UI widgets that are better tailored to different user preferences and task contexts. Although broadly applicable, we focus on the image editing domain within a Python-based environment to demonstrate the framework’s effectiveness. We constructed a preference library by collecting data from 50 participants to capture general users' preferences across three core dimensions of UI design: \textit{predictability}, \textit{efficiency}, and \textit{explorability}. This preference data is then integrated into the LLM’s UI generation process to augment LLM-based widget reasoning and code generation to produce UI widgets that align with user and task demands.

In a user study with 78 additional participants, we found that UI widgets generated using our framework were rated as more aligned with user preferences and task requirements compared to those generated by the LLM alone. These results highlight the potential of our framework to advance user-centered interface generation and demonstrate that integrating crowdsourced preferences can significantly enhance the effectiveness of LLM-based UI generation.

In summary, our main contributions are as follows:
\begin{itemize}[left=0pt]
    \item We present a framework that enhances LLM-based UI generation by incorporating crowdsourced user preferences to support task-aware and preference-aligned UI widget generation. We introduce the framework design and the methodology for crowdsourcing user preference data.
    \item We implement a system prototype for this framework in the image editing domain. This implementation delivers 720 user preferences crowdsourced from 50 users and a structured reasoning technique to guide LLM reasoning and code generation for user preference-aligned UI generation.
    \item We conduct a user study with an additional 78 participants that discovers the impact of crowdsourced user preference data on UI generation. By discussing the study findings, we provide insights for future research on creating adaptive and user-centered LLM-generated UIs with user data as guidance.
 \end{itemize}

\section{Related Work}
\label{sec:relatedwork}
To provide background for our work, we introduce prior research on traditional fixed UIs, natural language UIs, and dynamically generated UIs, and discuss their advantages as well as limitations. 

\subsection{Traditional Fixed UIs}
Traditional UI design in professional digital software like Illustrator~\cite{illustrator}, Photoshop~\cite{photoshop}, Premiere Pro~\cite{premierepro}, and Blender~\cite{blender} follows a designer-centric model, where a fixed set of tools and UI controls are designed for various tasks. These applications rely on well-established interface elements such as dropdown menus, sliders, text fields, and panels that allow users to manipulate images, graphics, videos, and 3D models. For example, Photoshop provides consistent tools for adjustments like brightness, contrast, and hue, while Illustrator offers vector editing through pre-defined tools. Premiere Pro delivers video editing features through timelines and control panels, and Blender offers a complex interface for 3D modeling and animation. These tools are powerful yet rigid, requiring users to learn and adapt to the software’s structure. As a result, their complexity can overwhelm beginner users, and their interface, though it has undergone interactive user testing, may not always align with individual workflows or preferences, limiting the flexibility and personalization of the user experience.

\subsection{Natural Language UIs}
Using natural language as input to edit digital content has gained increasing interest due to its simplicity. This approach allows users to perform tasks like the generation and editing of images~\cite{dalle, runaway, firefly, generativefill, brooks2023instructpix2pix}, videos~\cite{descript, premierepro_sensei, qin2024instructvid2vid}, data visualizations~\cite{maddigan2023chat2vis, sah2024generating, wang2023data, wang2024dataformulator2, dibia2023lida}, 3D models~\cite{chang2017sceneseer, poole2022dreamfusion}, and human motions~\cite{tevet2023human, raab2023single, chen2023executing, liu2024exploring, liu2024dancegen} by simply describing the outputs they want rather than navigating complex software interfaces. 
Natural language UIs are transforming traditional workflows by making advanced editing more accessible to non-expert users, reducing the technical execution gulf, and enabling users to focus on creative intent. However, these UIs require articulation and can hurt usability as half the population is insufficiently articulate in writing to use natural language UIs well~\cite{nielsen2025articulation}. Natural language UIs also struggle with precise control over specific details. They cannot effectively support interactive editing, which limits their ability to handle complex, fine-grained adjustments and provide prompt visual feedback of intermediate results typically required in iterative workflows.

\subsection{Dynamically Generated UIs} \label{sec:rw_dynamic_ui}
Recent research has studied generating dynamic UIs to provide users with precise, direct interactions for content editing. One prominent approach is the relaxation method, which involves creating UI widgets that provide generalized control over variables within a function or query. For instance, Mavo~\cite{verou2016mavo} allows users to construct interactive HTML pages by incorporating unique attributes, which, in turn, recommend editing widgets tailored to those attributes. Heer et al.~\cite{heer2008generalized} employ query relaxation techniques, allowing users to broaden their selection tasks effectively. Bespoke~\cite{vaithilingam2019bespoke} utilizes user demonstrations to generate custom UIs for command-line applications. It applies rule-based heuristics to infer the semantic types of parameters in bash commands and thus dynamically create widgets for parameter manipulation. 
Another line of research, known as precision interfaces, uses SQL queries as a framework for UI generation~\cite{zhang2018precision, chen2020monte, chen2022pi2, chen2022nl2interface}. These systems convert a sequence of input queries into interactive widgets, allowing users to adjust query parameters directly. 

In addition to generating UIs from command-line applications and SQL queries, recent advancements have focused on creating UIs from natural language. This approach allows simplicity by taking users' natural language as input, and the system generates an interactive UI tailored to users' task descriptions using LLM's reasoning and code generation capabilities. For instance, DynaVis~\cite{vaithilingam2024dynavis} generates dynamic interfaces for visualization editing based on natural language commands. By enabling users to describe changes to their visualizations through simple text, such as ``\textit{rotate the x-labels by 30 degrees counter-clockwise}'', DynaVis dynamically creates suitable widgets, such as a slider, for adjusting the rotation of x-labels. Similarly, Biscuit~\cite{cheng2024biscuit} generates UIs to help users follow machine learning tutorials in Jupyter Notebooks using natural language queries. Users can describe tasks like ``\textit{sample training data}'' or ``\textit{adjust model hyperparameters}'', and Biscuit generates UIs for the specified parameter adjustments. 
Although natural language-based UI generation supports both precision with interactable UI widgets and simplicity by taking natural language input, most prior work using standalone LLMs does not fully consider individual user preferences and domain-specific task requirements. This can result in generated interfaces that, while functional, may not be the most effective for specific users, especially when personalization and task-specific requirements are needed.

\begin{figure*}[!ht]
    \centering
    \includegraphics[trim=90 80 90 80, clip, width=0.7\textwidth]{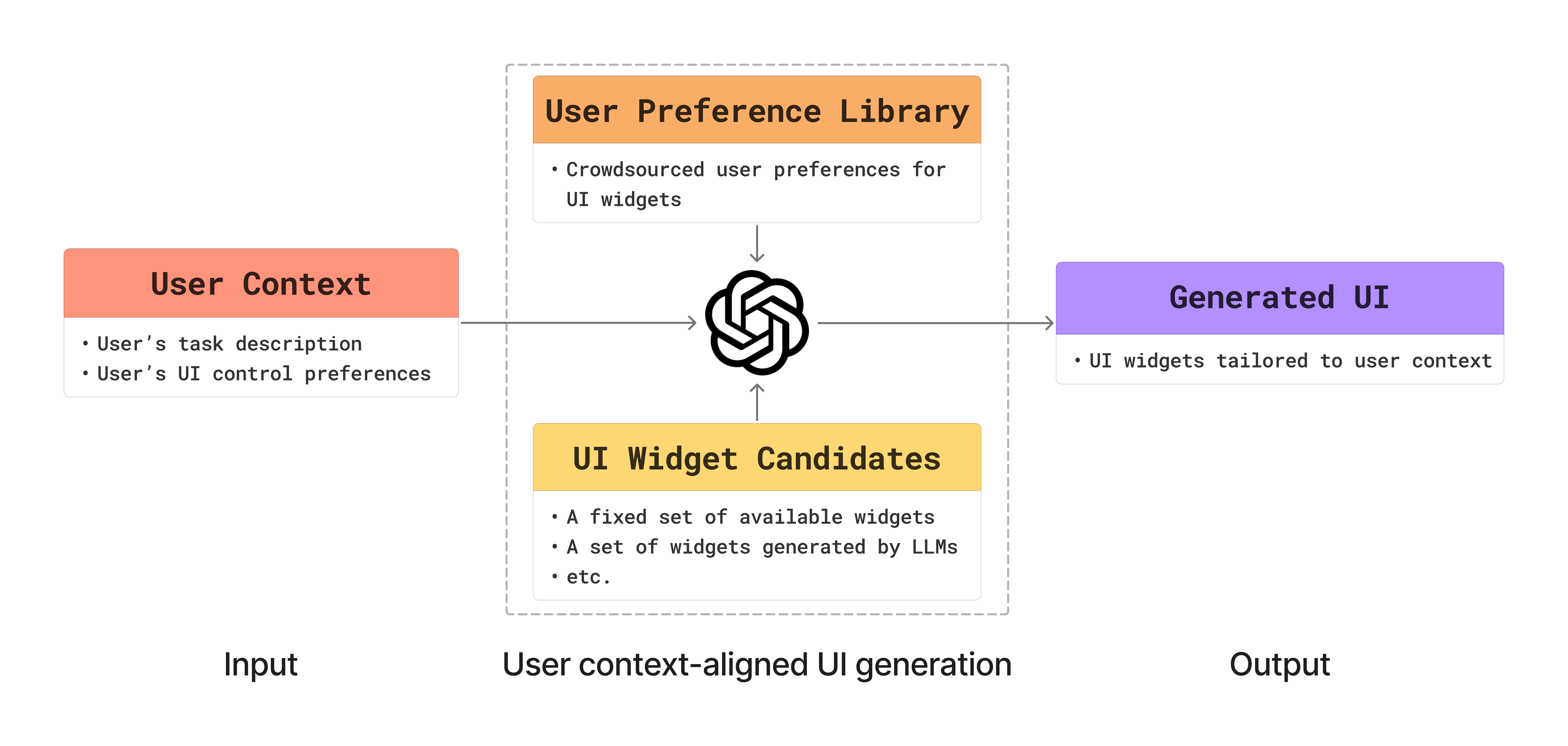}
    \caption{Overview of the \crowdgenui{} framework. This framework consists of four main components: it takes (1) user context as input, including task description and UI control preferences, and this information is passed to the LLM together with (2) a user preference library and (3) a collection of UI widget candidates. All the input data is used to guide the LLM reasoning process to obtain (4) generated UI widgets that are aligned with user preferences and task requirements.}
    \Description{This figure is a flow chart that takes user context, user preference library, and UI widget candidates as input for LLMs, and generates UI as the output.}
    \label{fig:framework_overview}
\end{figure*}

\section{\crowdgenui{}: A Framework that Aligns UI Generation with User Preference}
\subsection{Overview}
Figure~\ref{fig:framework_overview} provides an overview of our proposed framework, \crowdgenui{}, which aligns generated UI widgets with task requirements and specific user preferences by incorporating a user preference library that captures users' preferred UI widgets and their rationale to guide LLM-based UI generation. 

\textbf{Enhanced alignment through user preference integration}:
Unlike prior work on LLM-based UI generation, this framework integrates user preference insights directly into the LLM reasoning process. The user insights consist of crowdsourced preference data for UI widgets from real users. By designing a systematic reasoning technique to guide LLM reasoning, this framework is designed to boost LLM alignment in the UI generation application. 

\textbf{Flexibility and extendability through modular design}: 
By using a modular-based architecture, this framework allows independent and flexible module updates to accommodate extendable contexts. This applies to all three input modules: (1) user context can evolve with updated task requirements and various preferences, ensuring continual alignment with specific user needs; (2) user preference library can continuously take updated user data, expanding the diversity and richness of user insights used during UI generation; (3) UI widget candidates can flexibly incorporate either static widget sets from established libraries or dynamically produced widgets, thereby expanding potential UI outcomes. 

\subsection{Usage Scenario}
Before diving into the technical implementation details, we guide readers through an example usage scenario, shown in Figure~\ref{fig:usage_scenario}, to demonstrate how \crowdgenui{} can be used for a content editing task. Imagine that Alice wants to change the tone of an image from green, spring colors to warm, fall colors. She uses \crowdgenui{} to edit her image through the following process:

\begin{figure*}[!ht]
    \centering
    \includegraphics[trim=90 80 90 80, clip, width=0.8\textwidth]{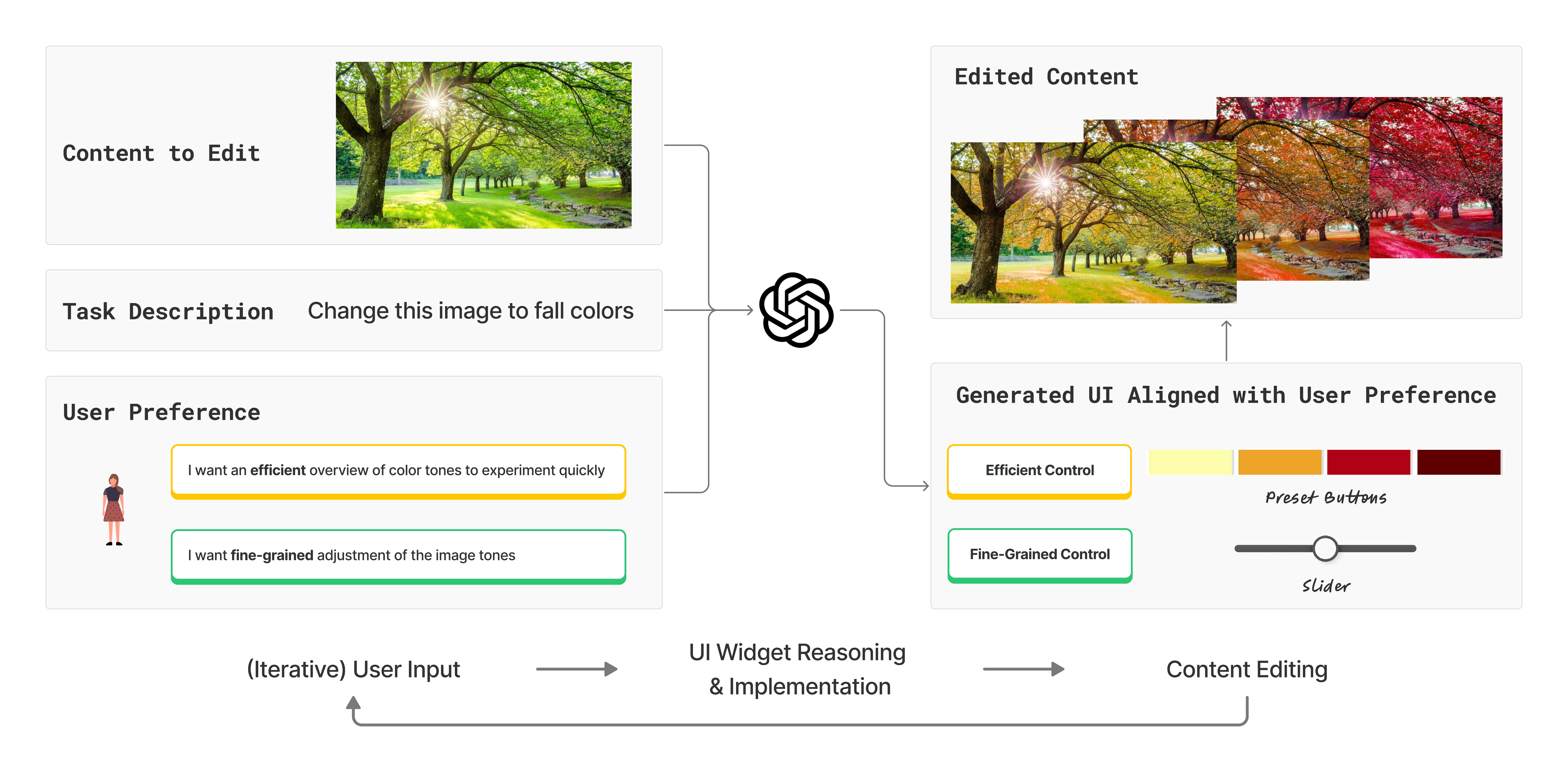}
    \caption{Example usage scenario of \crowdgenui{}. A user wants to change the color of an image to fall colors. They provide the image, describe the task, and specify their UI control preference. Taking user input, \crowdgenui{} reasons the proper UI widgets and generates the code for widget implementation. The user uses the generated widgets to edit the image and can further update their input to obtain updated UIs, e.g., from efficient experiments using preset options to fine-grained control using a slider.}
    \Description{This figure demonstrates an example usage scenario using a flow chart. From left to right, the chart consists of interactive user input, UI widget reasoning and implementation, and content editing.}
    \label{fig:usage_scenario}
\end{figure*}

\textbf{Step 1---Describe task and specify preferences}: She describes her needs as ``\textit{Change this image to fall colors}'' and provides the image she wants to edit. She also specifies that she prefers to experiment with image colors quickly with color tone overviews.

\textbf{Step 2---Use \crowdgenui{} for UI widget reasoning and implementation}: \crowdgenui{} processes Alice’s input and determines the most suitable UI widgets for her task. Since the task involves adjusting the image’s color tone efficiently, \crowdgenui{} consults the crowdsourced user preference library to identify relevant tasks and commonly preferred UI widgets for efficiency. Based on this information, it reasons that preset buttons displaying color tone previews are the optimal choice. Next, it generates code to implement an image tone adjustment function and a preset button function, links the preset buttons to image tone adjustment, and returns the code for Alice to execute and obtain the widgets.

\textbf{Step 3---Edit content with generated UI}: Alice interacts with the preset buttons to explore various pre-defined tones. The color previews on buttons help her to anticipate how each selection will affect the image, and the preset values enable quick experimentation with simple button clicks. 

\textbf{Step 4---Update requirements for iterative UI generation}: Although the preset buttons satisfy Alice's initial needs, she finds an orange tone she likes but wants to make it slightly redder. Since the preset options do not allow that, she updates her request by asking for UI widgets that support finer tone adjustments. This brings her back to Step 2, where \crowdgenui{} generates new widgets based on her updated requirements, and moves on to Step 3, where she continues refining her image using the newly generated widgets.

\section{\crowdgenui{} Implementation}
\label{sec:system}
\begin{figure*}[!ht]
    \centering
    \includegraphics[trim=90 80 90 80, clip, width=\textwidth]{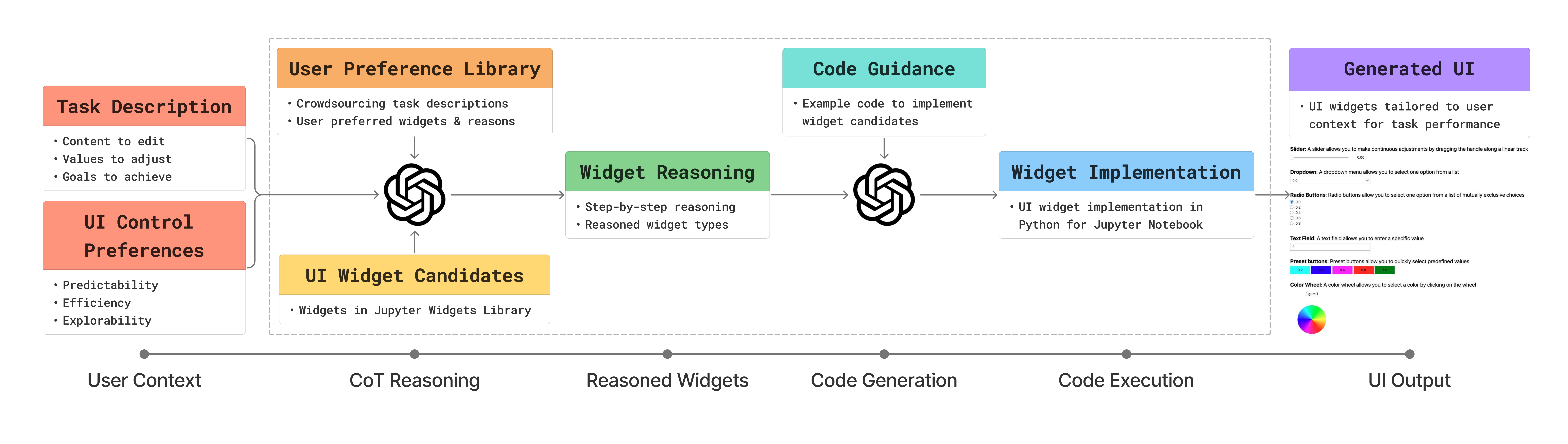}
    \caption{System implementation pipeline. The system starts with natural language input specifying the task description and user preferences. With this input, the LLM applies chain-of-thought (CoT) reasoning to obtain reasoned widgets. Based on this reasoning process, the crowdsourced user preference library, containing the crowdsourcing tasks, user-preferred widgets, and their rationale, is integrated to augment widget reasoning. Additionally, the Jupyter Widget Library widget candidates are fed to the LLM for reasoned widget implementation. Lastly, the reasoned widgets and reference code snippets are passed to the LLM for code generation in Python. This allows widget implementation and interaction in Jupyter Notebook.}
    \Description{This figure is a flow chart showing the system implementation pipeline. Starting from the left, the task description, preference aspect, crowdsourced preference library, and UI widget candidates are passed as input to the LLM for CoT reasoning. As the reasoning output, reasoned widgets containing widget types and the corresponding rationale are returned. This output is further passed to the LLM for code generation with sample code as guidance. Finally, Python code is returned for widget display and interaction.}
    \label{fig:llm_ui_generation_pipeline}
\end{figure*}

Figure~\ref{fig:llm_ui_generation_pipeline} illustrates a system prototype we implemented for our proposed framework. In this section, we dive into the technical details and implementation rationale.

\subsection{User Context}
The system input is user context, which consists of task description and UI control preferences. 

\subsubsection{Task Description}
The task description needs to specify the content to edit (e.g., an image), values to adjust (e.g., image color), and goals to achieve (e.g., changing the image to fall colors). 

\subsubsection{UI Control Preferences} \label{sec:user_preference_aspects}
UI control preferences can involve various preference aspects. In the current system, we use \textit{predictability}, \textit{efficiency}, and \textit{explorability} to implement our prototype. 

These preferences are chosen based on well-established UI design principles developed by prior research. Specifically, McKay's \textit{UI is Communication}~\cite{mckay2013ui} emphasizes that an intuitive interface relies on effective communication that balances multiple principles, including predictability, efficiency, discoverability, and explorability. McKay argues that intuitive UIs align with a user's prior experiences, providing clear affordances and responsive feedback to guide interaction. Lidwell et al.'s \textit{Universal Principles of Design}~\cite{lidwell2010universal} expands on these ideas by highlighting fundamental design principles such as the flexibility-usability tradeoff, mapping, and mental models, which underscore the importance of aligning system controls with user expectations and simplifying complex interactions. Together, these principles highlight that successful UIs are predictable and efficient and allow for exploratory interaction without overwhelming the user, leading to our selection of \textit{predictability}, \textit{efficiency}, and \textit{explorability} to guide LLM-based UI generation. 
We introduce the user preference library that we crowdsourced according to the three aspects in Section~\ref{sec:crowdsourcing}.

\subsubsection{Input to the LLM}
Apart from feeding the image to the LLM, the rest of the user context is represented as text input denoted by task-description ($T$) and preference-aspect ($A$) in JSON format. 

\subsection{UI Widget Reasoning}
User context is passed to the LLM for widget reasoning. Additional input for LLM reasoning includes the crowdsourced user preference library and UI widget candidates. Crowdsourced-library ($L$) and widget-candidates ($C$) are also text data formatted as JSON. With all the input data, we instruct the LLM to use a chain-of-thought (CoT) reasoning technique to reason UI widgets. We present the CoT reasoning overview in Algorithm~\ref{algo:cot_reasoning} and the detailed prompt in Appendix~\ref{appendix:prompt_widget_reasoning_withlib}.

The LLM first analyzes the user's task description to identify the task requirements. Based on this analysis, it is instructed to analyze the similarity between the given task and the crowdsourcing tasks by comparing task descriptions to extract similar tasks. 
Once relevant tasks are identified, the LLM is further prompted to look for user-preferred widgets for each relevant task according to the preference aspect and the reasons users provided for their choices. It needs to refer to the located crowdsourced data to reason the most appropriate widgets that belong to the widget candidates.
Finally, the UI widget reasoning results are formatted as JSON, including the user task's name, the reasoned widgets for each preference aspect, and the corresponding reasoning.

\begin{algorithm}
\caption{Overview of CoT Reasoning}
\label{algo:cot_reasoning}

\begin{algorithmic}[0]
\STATE \textbf{Input}: task-description ($T$), preference-aspect ($A$), crowdsourced-library ($L$), widget-candidates ($C$)
\STATE \textbf{Output}: widget-reasoning
\end{algorithmic}

\begin{algorithmic}[1]
\FOR{each crowdsourced-task in $L$}
    \STATE relevant-tasks $\gets$ \textit{similarity} ($T$, crowdsourced-task) \COMMENT{Identify relevant tasks}
    
    \FOR{each aspect in $A$}
        \STATE ui-widgets $\gets$ \textit{match} (aspect, relevant-task-aspect) \COMMENT{Find relevant widgets}
        \RETURN ui-widgets $\in$ $C$ \COMMENT{Return matching widgets}
    \ENDFOR
\ENDFOR

\STATE widget-reasoning $\gets$ JSON (ui-widgets) \COMMENT{Convert results to JSON format}
\end{algorithmic}
\end{algorithm}

\subsection{UI Widget Implementation}
Once the reasoned widgets are obtained, the LLM is further used to generate the Python code to implement these widgets. During this process, we provide an example code, which covers the implementation of all the widget candidates, as guidance for the LLM to reference during code generation. Our experiment shows that including example code significantly reduces bugs in the LLM-generated code. More importantly, the example code enables the LLM to successfully generate widget implementations that go beyond the default functionality of the Jupyter Widgets Library. For instance, a color wheel widget requires additional use of the Interactive Canvas in Jupyter~\cite{ipycanvas} and the implementation of mouse interactions on the wheel to map clicks to the corresponding colors. In addition to specific widget implementations, the example code also offers a structure to organize and display widgets in a clear, structured format within Jupyter Notebook for user interaction. The detailed prompt and example code for UI widget implementation can be found in Appendix~\ref{appendix:prompt_widget_coding}. 

\subsection{System Implementation Rationale}
\label{sec:implementation_rationale}

\textbf{Task domain}:
We implement the system in the image editing domain because it offers generalizability across a diverse range of content editing domains. We carefully design image editing tasks that involve various data manipulation needs, including continuous and discrete value adjustments and tasks emphasizing color and position visual feedback (Section~\ref{sec:crowdsourcing_tasks}). These tasks require both precise control and iterative exploration, which are essential for content editing requirements in general. Moreover, the interactions employed in image editing, characterized by real-time visual feedback, trial and error, and fine-tuning content attributes, are directly transferable to other digital content editing applications such as video editing, data visualization, and graphic design. Consequently, image editing tasks address fundamental data manipulation needs that apply to broader digital content editing practices.

\textbf{User interface}:
The current implementation uses Jupyter Notebook to build the user interface as it can handle all the data manipulation requirements for our tasks. Although this implementation does not cover other UI possibilities, such as web or professional software, the framework is extendable to accommodate these UIs by updating the widget candidates to web or software widget libraries and generating code in the corresponding programming languages for widget implementation. 

\textbf{UI widget candidates}:
The current implementation focuses on the Jupyter Widgets Library widgets rather than open-ended options. This allows practical feasibility since implementing an infinite variety of widgets can be unmanageable. Also, these widgets are strategically chosen as foundational elements that map to universal interaction principles. For instance, sliders and text fields for continuous adjustments can extend to more granular controls like range sliders or numeric steppers, and dropdowns and radio buttons for discrete selections can evolve into multi-select interfaces or dynamic filters. Focusing on these core widgets can allow scalability by integration with external libraries (e.g., Angular~\cite{angular}, React Bootstrap~\cite{react_bootstrap}, or Vuetify~\cite{vuetify}) to expand into advanced UI components in future iterations.

\textbf{LLM selection}:
We use GPT-4o~\cite{gpt4o} for UI widget reasoning and implementation due to its advanced reasoning capabilities, robust code generation performance, and strong alignment with natural language instructions, which are critical for dynamically generating UIs based on task descriptions and user preferences.

\section{User Preference Crowdsourcing}
\label{sec:crowdsourcing}
In this section, we introduce our collected user preference data to build the library for UI generation. We present why we use crowdsourcing to collect this data, followed by the crowdsourcing setup and an analysis of the collected data.

\begin{figure*}[!ht]
    \centering
    \includegraphics[trim=90 80 90 80, clip, width=0.85\textwidth]{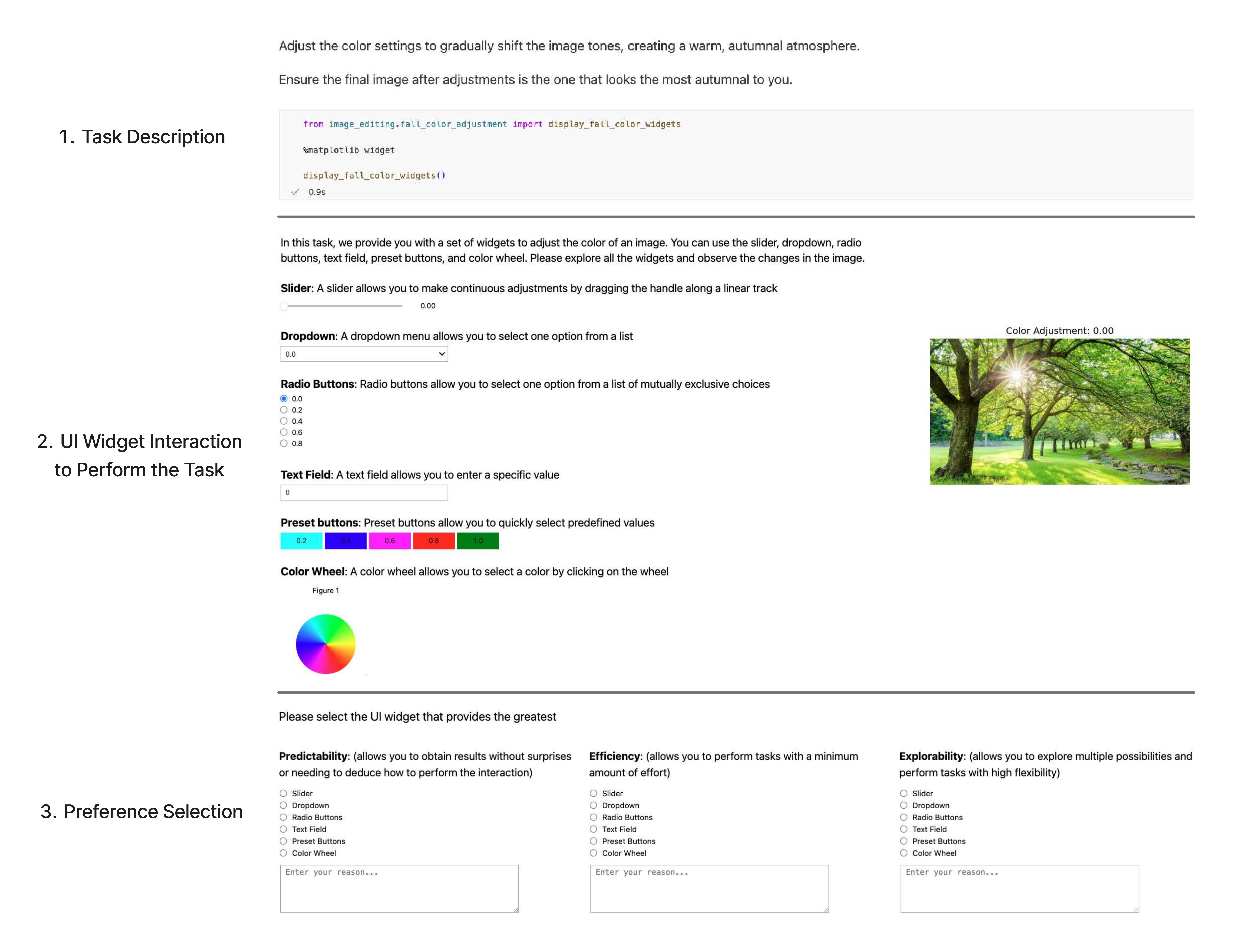}
    \caption{Crowdsourcing user interface for task: adjusting an image to fall colors (\texttt{image\_adjust\_fall\_color}).}
    \Description{This figure shows an example of the crowdsourcing user interface consisting of three components. From top to bottom, the components are the task description, the UI widget interaction panel to perform the task, and the user preference selection panel. The task description specifies the goal that the user is expected to achieve. Below it, the UI widget interaction panel shows multiple widgets on the left and an image to edit on the right. At the bottom, the preference selection panel allows users to select their preferred widgets and write their reasons for the three preference aspects.}
    \label{fig:ui_image_adjust_fall_color}
\end{figure*}

\subsection{Motivation}
User preference is part of user feedback and plays a crucial role in UI design and UI widget selection. The user feedback data is traditionally collected through labor-intensive methods like interviews, usability tests, A/B tests, and feedback forms~\cite{stone2005user, wilson2013user}. Given that LLMs are trained on large-scale human data, they provide a promising solution to incorporating user feedback, such as user preferences, into dynamically generated UIs. However, without dedicated user preference datasets, current LLMs cannot fully address this need (recall Section~\ref{sec:rw_dynamic_ui}). To fill this gap, we propose crowdsourcing a preference library to enhance LLM-based UI widget generation with user preference data, motivated by the effectiveness of adopting crowdsourcing for user feedback collection to assist various design areas in previous research endeavors~\cite{xu2014voyant, hossain2015crowdsourcing, xu2015classroom, luther2015structuring, yu2016encouraging, foong2017novice, lee2018exploring, oppenlaender2020crowdui}.

\subsection{Participants}
The crowdsourcing took place on Prolific~\cite{prolific}. We recruited participants with Python programming experience only to ensure they were familiar with running the provided Python code in Jupyter Notebook. During the study, they did not write code but just executed our provided code to interact with UI widgets and work on their tasks on desktops using a mouse and keyboard. 
We did not require participants to have image editing or UI programming experience, as we aim to collect regular users' preference data and make generated UIs align with end users' preferences. 

We acknowledge that this participant recruitment may limit our sample by excluding users who are unfamiliar with Python. However, these participants can still meaningfully contribute to general users' UI preferences, as programming expertise is independent of UI usage and preference for non-programming tasks. 

According to prior research on user testing for UX design~\cite{tomczak2023over, usertesting2025sample, userinterviews2025preference}, 20–30 participants are generally sufficient for evaluating simple to intermediate tasks, while 50 participants are recommended for more complex tasks. Following these established guidelines, we invited 50 participants to the crowdsourcing study (Age: 29.86 $\pm$ 7.83, Male: 37, Female: 12, Non-binary: 1), and compensated them 10 USD for the 30-minute study. 

\subsection{Crowdsourcing Setup}
\subsubsection{User Interface}
During the crowdsourcing study, each participant had a Jupyter Notebook to work on eight image editing tasks, interact with multiple UI widgets to perform each task, and provide their preferences for the widgets. Figure~\ref{fig:ui_image_adjust_fall_color} shows the UI for a crowdsourcing task---adjusting an image to fall colors. The interface contains the task requirement, multiple UI widgets, an image to edit, and a user preference selection panel. 

\subsubsection{Tasks}
\label{sec:crowdsourcing_tasks}
The crowdsourcing tasks cover three categories in the image editing domain. Some tasks may overlap across multiple categories. Our goal is to use these tasks to incorporate a range of value adjustments and data manipulation needs.

\textbf{Continuous value adjustment}: These tasks need to adjust a value across a continuous range. Examples include adjusting image lightness, saturation, hue, and color balance. 

\textbf{Discrete value selection}: These tasks involve selecting from predefined options. Examples include choosing predefined values for image lightness, saturation, hue, and watermark positions.

\textbf{Tasks emphasizing visual feedback}: These tasks require immediate visual feedback, such as color or position changes. Examples include changing an image to fall tones, matching an object's color to a given color, positioning a watermark, and highlighting a specific image area.

The detailed task descriptions are presented in Table~\ref{tab:crowdsourcing_tasks}. We use these tasks to encourage exploration, such as in Tasks 1, 2, 3, 6, and 7, while also emphasizing precision, as seen in Tasks 4, 5, and 8. Tasks 4 and 5 are used as control tasks for attention checks to help us identify untrustworthy responses, where we instruct participants to ensure that the final images best meet the task requirements. We randomize the order of tasks, ensuring that no particular task appears consistently in the same position to reduce order effects that could influence user behaviors. 

\begin{table}[!ht]
    \centering
    \caption{Crowdsourcing task names and descriptions.}
    \Description{This table consists of two columns: task name and task description. In each row, the number, name, and description of each crowdsourcing task are detailed.}
    \scalebox{0.8} {
    \begin{tabular}{cp{0.35\textwidth}p{0.65\textwidth}}
    \toprule
         & \multicolumn{0}{c}{\textbf{Task Name}} & \multicolumn{0}{c}{\textbf{Task Description}}\\ \midrule
        1 & \texttt{image\_adjust\_lightness} & Experiment with lightness settings to see how different levels affect the image.\\ \hline
        2 & \texttt{image\_adjust\_saturation} & Boost the saturation to make the colors pop.\\ \hline
        3 & \texttt{image\_adjust\_hue} & Experiment with different hues to find a color tone that complements the overall mood of the image.\\ \hline
        4 & \texttt{image\_adjust\_fall\_color} & Adjust the color settings to gradually shift the image tones, creating a warm, autumnal atmosphere. Ensure the final image is the one that looks the most autumnal to you.\\ \hline
        5 & \texttt{image\_color\_match} & Change the rocket's color to match the provided reference color. Ensure the rocket's color in the final image best matches the reference color.\\ \hline
        6 & \texttt{image\_adjust\_color\_balance} & Experiment with different color balance settings to see how altering the red, green, and blue levels affects the overall color harmony of the image.\\ \hline
        7 & \texttt{image\_place\_watermark} & Experiment with different watermark positions to find the perfect balance between visibility and subtlety.\\ \hline
        8 & \texttt{image\_place\_vignette} & Darken the background using a vignette effect except for the human face by properly positioning the circle around the face.\\
    \bottomrule
    \end{tabular}
    }
    \label{tab:crowdsourcing_tasks}
\end{table}

\subsubsection{User Preference Aspects}
We use predictability, efficiency, and explorability as user preference aspects, drawing from prior research on intuitive user interface design principles (recall Section~\ref{sec:user_preference_aspects}). (1) Predictability emphasizes that users can achieve results without encountering unexpected outcomes or needing to figure out how to perform the interaction. (2) Efficiency highlights that users should be allowed to perform tasks with minimal effort. (3) Explorability underscores that users can explore multiple possibilities and perform tasks with high flexibility.

\subsubsection{UI Widget Candidates}
\label{sec:widget_selection_implementation}
As presented in Section~\ref{sec:implementation_rationale}, we aim to select UI widgets to cover various data manipulation requirements and align these candidates with the specified task requirements. To this end, the Jupyter Widgets Library~\cite{ipywidgets} offers both basic and advanced UI widgets, including a wide range of control needs. Specifically, for continuous value adjustment tasks, sliders are selected to allow for fine-grained control, and text fields can be used for precise value input; for discrete value selection tasks, dropdown menus and radio buttons are often effective; for tasks emphasizing visual feedback, UI widgets like color pickers, direct manipulation (e.g., clicking), and preset buttons with preview overlays are picked to show outcome previews through color or position cues.

\subsection{Crowdsourced UI Preference Library}
\label{sec:crowdsourcing_result_analysis}
\subsubsection{Statistics Overview}
\begin{figure*}[!ht]
    \centering
    \includegraphics[width=0.8\textwidth]{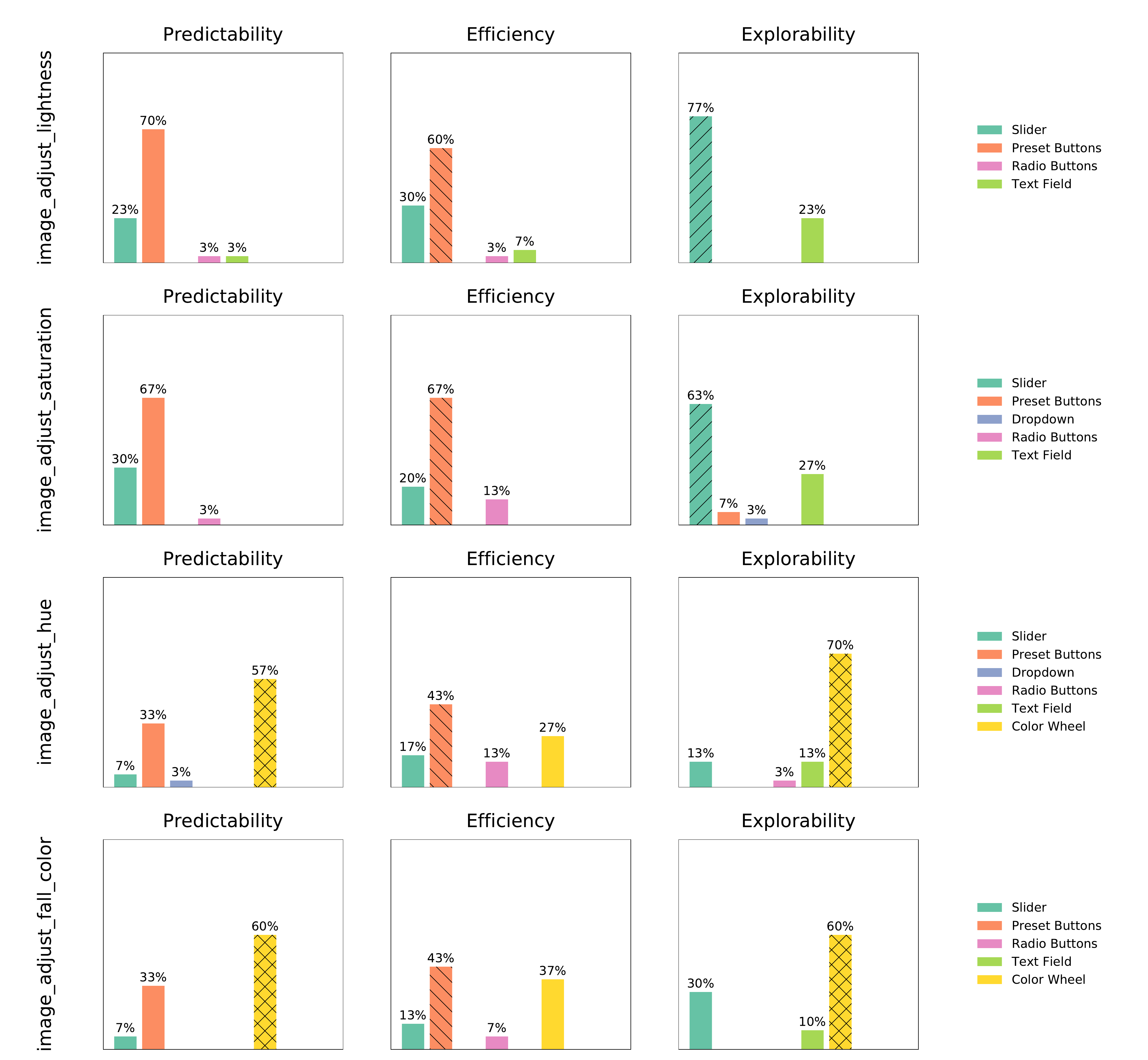}
    \caption{User preference distribution of tasks 1-4 (task set 1) collected in the crowdsourcing study.}
    \Description{This figure consists of four rows, showing the crowdsourcing results of tasks 1 to 4 in each row. Each row has three columns, depicting the user preference distribution of the three preference aspects in each column as bar charts.}
    \label{fig:crowdsourcing_preferences_1to4}
\end{figure*}

\begin{figure*}[!ht]
    \centering
    \includegraphics[width=0.8\textwidth]{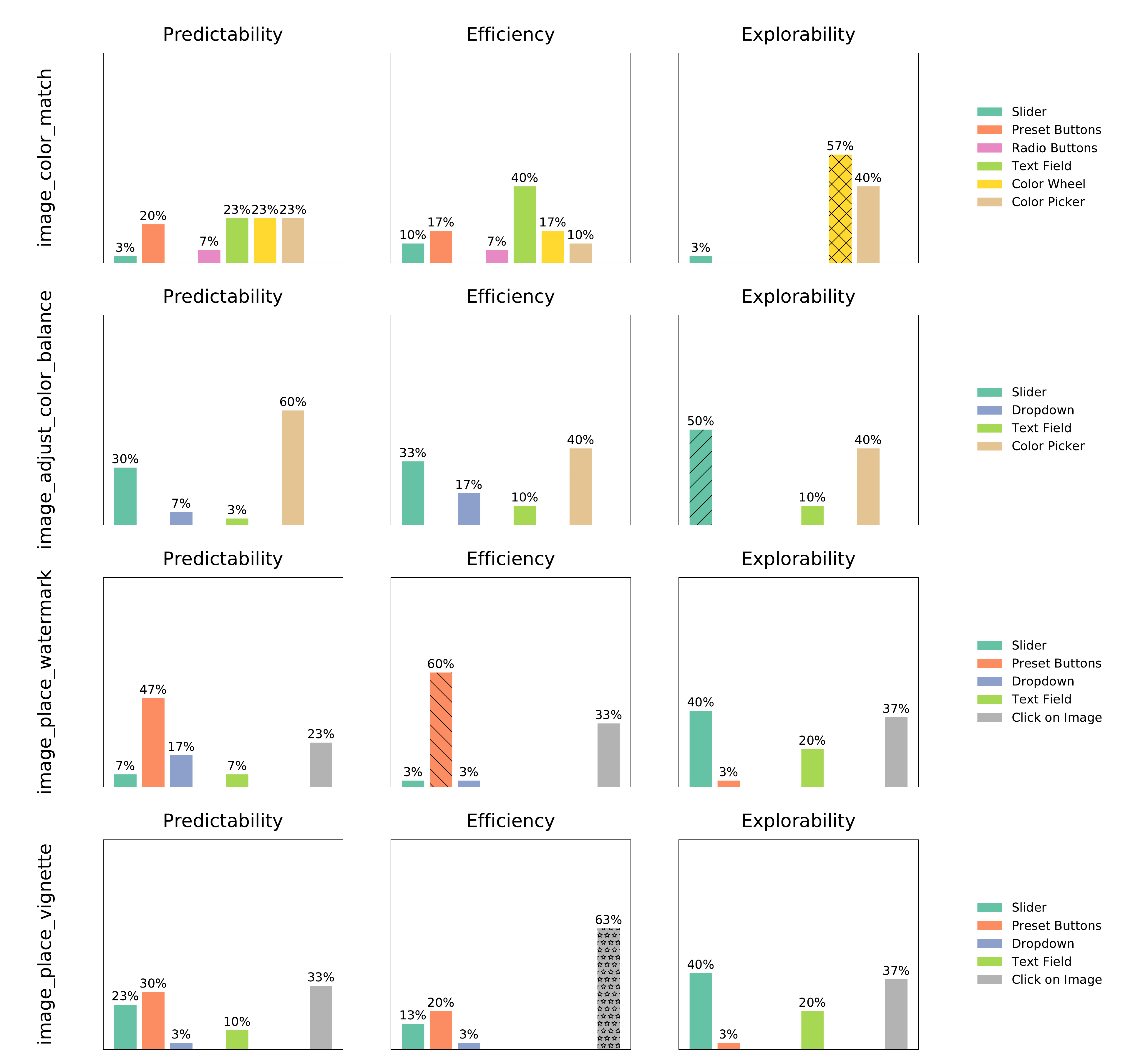}
    \caption{User preference distribution of tasks 5-8 (task set 2) collected in the crowdsourcing study.}
    \Description{This figure consists of four rows, showing the crowdsourcing results of tasks 5 to 8 in each row. Each row has three columns, depicting the user preference distribution of the three preference aspects in each column as bar charts.}
    \label{fig:crowdsourcing_preferences_5to8}
\end{figure*}

We distributed the crowdsourcing study to our participants in three versions. Each version included all eight tasks and a subset of the three preference aspects, covering one, two, or all three aspects. This approach ensured that each participant provided their preferences for 15-18 preference selections out of the total 3 aspects $\times$ 8 tasks = 24 selections across all tasks, helping to prevent fatigue. The study yielded a UI widget preference library with 30 unique user responses for each aspect. In total, we obtained 30 responses $\times$ 3 aspects $\times$ 8 tasks = 720 user preferences. Figures~\ref{fig:crowdsourcing_preferences_1to4} and \ref{fig:crowdsourcing_preferences_5to8} show the distribution of collected user preferences.

\subsubsection{Analysis of User Preferences}
We observe variations in user preferences for UI widgets regarding predictability, efficiency, and explorability across the eight tasks.

\textbf{Continuous value adjustment}: Users preferred preset buttons with preview overlays for their predictability and efficiency. For explorability, sliders were favored for adjusting lightness and saturation, while a color wheel was preferred to explore hue adjustment.

\textbf{Discrete value selection}: Preset buttons were generally favored over radio buttons and dropdown menus in all three preference aspects, primarily due to the visual cues provided.

\textbf{Tasks emphasizing visual feedback}: User preferences varied based on whether the task involved color or position adjustments. (1) In color adjustment tasks emphasizing exploration, many users preferred color wheels and color pickers for predictability and explorability, while preset buttons were selected for efficiency, similar to continuous and discrete value adjustment tasks. (2) However, for tasks requiring precise color adjustment, such as matching an object’s color to a given color, there was no clear consensus regarding predictability. Participants were evenly split among text fields, color wheels, color pickers, and preset buttons. Regarding efficiency, most users preferred text fields. For explorability, color wheels and color pickers were the most favored. 
(3) On the other hand, in position adjustment tasks, preset buttons and directly clicking on the image were preferred for predictability and efficiency, while for explorability, users favored both clicking on the image and using sliders.

\section{User Evaluation}
\label{sec:evaluation}
In this section, we introduce the user evaluation to assess our implemented system for the \crowdgenui{} framework. We outline the research questions to evaluate in the study, describe the evaluation setup, and present the evaluation results.

\subsection{Research Questions} \label{sec:research_questions}
We aim to study the following research questions in our user evaluation, focusing on the effectiveness and generalizability of the proposed framework and the impact of integrating the user preference library on UI generation:

\textbf{RQ1---Adaptivity of the framework to new, related tasks}: Does the framework enable the generalization of crowdsourced library-enhanced widget reasoning and generation to perform tasks that differ from, but relate to, the original crowdsourcing tasks?

\textbf{RQ2---Impact of the library on aligning generated widgets with user preferences}: Can the crowdsourced UI widget preference library help LLM-generated widgets align more closely with user preferences for predictability, efficiency, and explorability compared to those generated without a library?

\textbf{RQ3---Effect of library size on enhancing LLM reasoning}: Does the size of the crowdsourced UI widget preference library affect LLM widget reasoning effectiveness? Which size of the library provides the strongest support for reasoning capabilities?

In the following subsections, we explain the rationale of these RQs, introduce the user evaluation setup designed to answer them, and analyze our evaluation results in alignment with them.

\subsection{Participants}
We conducted the user evaluation on Prolific~\cite{prolific} and employed the same recruitment criteria as those in the crowdsourcing study to ensure consistency in participant background. The participants worked on the user evaluation tasks on desktops using a mouse and keyboard. In total, we recruited 78 participants (Age: 29.81 $\pm$ 9.09, Male: 56, Female: 19, Non-binary: 3) and compensated each participant 10 USD for the 30-minute study. 

\subsection{Evaluation Setup}
\subsubsection{Procedure}
The user evaluation invited voluntary participants, who were first presented with a consent declaration at the beginning of the study. Once participants agreed to participate, they accessed the study interface via a URL linked to our server. Before starting the tasks, they were given a briefing that outlined the study requirements. Full details of the user evaluation briefing and instructions can be found in Appendix~\ref{appendix:user_eval_briefing_instructions}. Following that, they worked on three tasks and answered six questions for each task to provide their feedback on LLM-generated widgets.

\subsubsection{User Interface}
\begin{figure*}[!ht]
    \centering
    \includegraphics[trim=90 80 90 80, clip, width=0.85\textwidth]{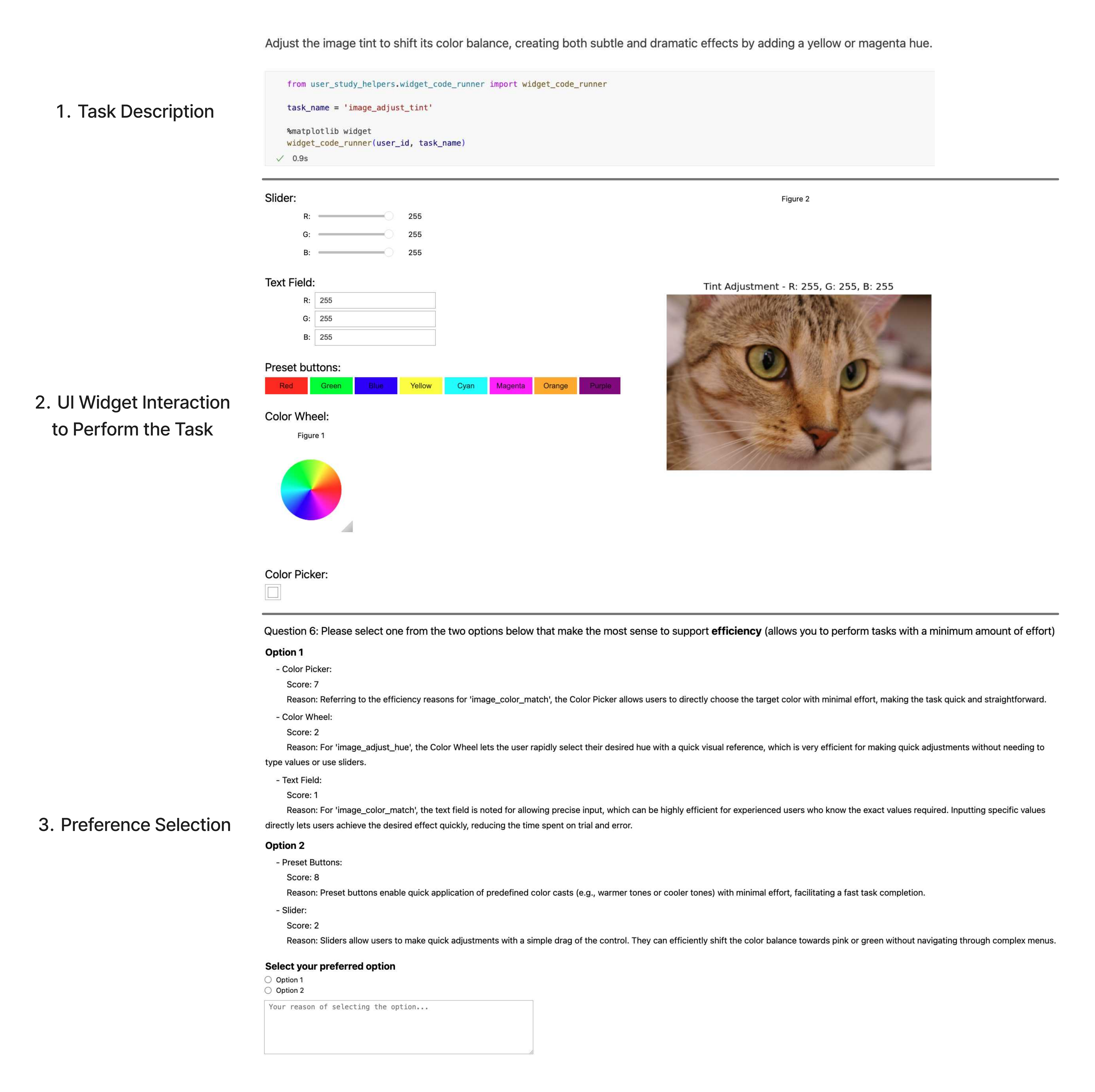}
    \caption{User evaluation user interface for task: adjusting image tint (\texttt{image\_adjust\_tint}). The preference selection panel contains six questions for each task, and we show one of them due to the space limit.}
    \Description{This figure shows an example of the user evaluation UI consisting of three components. From top to bottom, the components are the task description, the UI widget interaction panel to perform the task, and the user preference selection panel. The task description specifies the goal that the user is expected to achieve. Below it, the UI widget interaction panel shows multiple widgets on the left and an image to edit on the right. At the bottom, the preference selection panel allows users to select their preferred widget option and write their reasons for a certain preference aspect.}
    \label{fig:ui_image_adjust_tint}
\end{figure*}

Figure~\ref{fig:ui_image_adjust_tint} shows the user interface for a task in the user evaluation---adjusting image tint. Each participant interacted with a Jupyter Notebook containing LLM-generated UI widgets to work on tasks distinct from those in the crowdsourcing study. The UI widgets were generated by the LLM alone and enhanced by our crowdsourced library. After interacting with all the offered widgets, participants were shown a preference selection panel where they chose their favored options for each task.

\subsubsection{Tasks}
\label{sec:user_eval_tasks_aspects}
The user study tasks differ from the crowdsourcing tasks but still fall within the same three categories (recall Section~\ref{sec:crowdsourcing_tasks}). The detailed task descriptions are in Table~\ref{tab:user_eval_tasks}. 

\textbf{Continuous value adjustment}: Tasks include adjusting image exposure, tint, and temperature. 

\textbf{Discrete value selection}: Tasks include choosing predefined values for image exposure, tint, temperature adjustment, text alignment, and logo positioning.

\textbf{Tasks emphasizing visual feedback}: Tasks include changing an image’s color to spring colors, aligning text with the margin, and placing a logo.

\begin{table}[!ht]
    \centering
    \caption{User evaluation task names and descriptions.}
    \Description{This table consists of two columns: task name and task description. In each row, the number, name, and description of each user evaluation task are detailed.}
    \scalebox{0.8} {
    \begin{tabular}{cp{0.35\textwidth}p{0.65\textwidth}}
    \toprule
         & \multicolumn{0}{c}{\textbf{Task Name}} & \multicolumn{0}{c}{\textbf{Task Description}}\\ \midrule
        1 & \texttt{image\_adjust\_exposure} & Adjust the image exposure by both decreasing and increasing it. Find the exposure level that appears the most natural and visually pleasing to you.\\ \hline
        2 & \texttt{image\_adjust\_tint} & Adjust the image tint to shift its color balance, creating both subtle and dramatic effects by adding a yellow or magenta hue.\\ \hline
        3 & \texttt{image\_adjust\_temperature} & Adjust the image temperature to warm and cool tones. Experiment with different settings to achieve the most desired effect for you.\\ \hline
        4 & \texttt{image\_change\_to\_spring} & Transform the image to reflect vibrant spring colors, adding fresh, lively hues.\\ \hline
        5 & \texttt{design\_align\_text} & Align the text ``Poster Title'' perfectly along one of the margins. Experiment with various positions to achieve a balanced and visually pleasing design.\\ \hline
        6 & \texttt{design\_position\_logo} & Experiment with various placements for the logo to determine the most visually appealing position that enhances visibility and complements the overall design of the image.\\
    \bottomrule
    \end{tabular}
    }
    \label{tab:user_eval_tasks}
\end{table}

These tasks are designed to probe the framework’s generalizability beyond the original crowdsourcing scope to address \textbf{RQ1}. First, the tasks aim to evaluate the framework’s ability to reason UI widgets for nonoverlapping tasks involving data manipulation requirements varying from continuous and discrete value adjustments to tasks emphasizing visual feedback. Second, the inclusion of design tasks (e.g., text alignment and logo positioning) seeks to test cross-domain applicability, determining whether the system can extend beyond image editing despite the crowdsourced data’s focus on this domain. Third, the tasks are structured to balance precision with open-ended exploration, ranging from objective adjustments (e.g., margin alignment) to subjective, creativity-driven goals (e.g., change to spring colors). By replicating the original task categories while introducing novel domains and nuanced objectives, the study aims to evaluate if the framework can generalize to unseen tasks without domain-specific overfitting.

\subsubsection{User Preference Aspects}
The preference aspects, predictability, efficiency, and explorability, are drawn from the crowdsourcing study. We use the same aspects to evaluate whether user preferences collected from crowdsourcing can effectively lead to user preference-aligned widget generation for new tasks in the user evaluation. This setup addresses \textbf{RQ2}. 

\subsubsection{UI Widgets Generated with Different Library Sizes}
\begin{figure*}[!ht]
    \centering
    \includegraphics[width=0.85\textwidth]{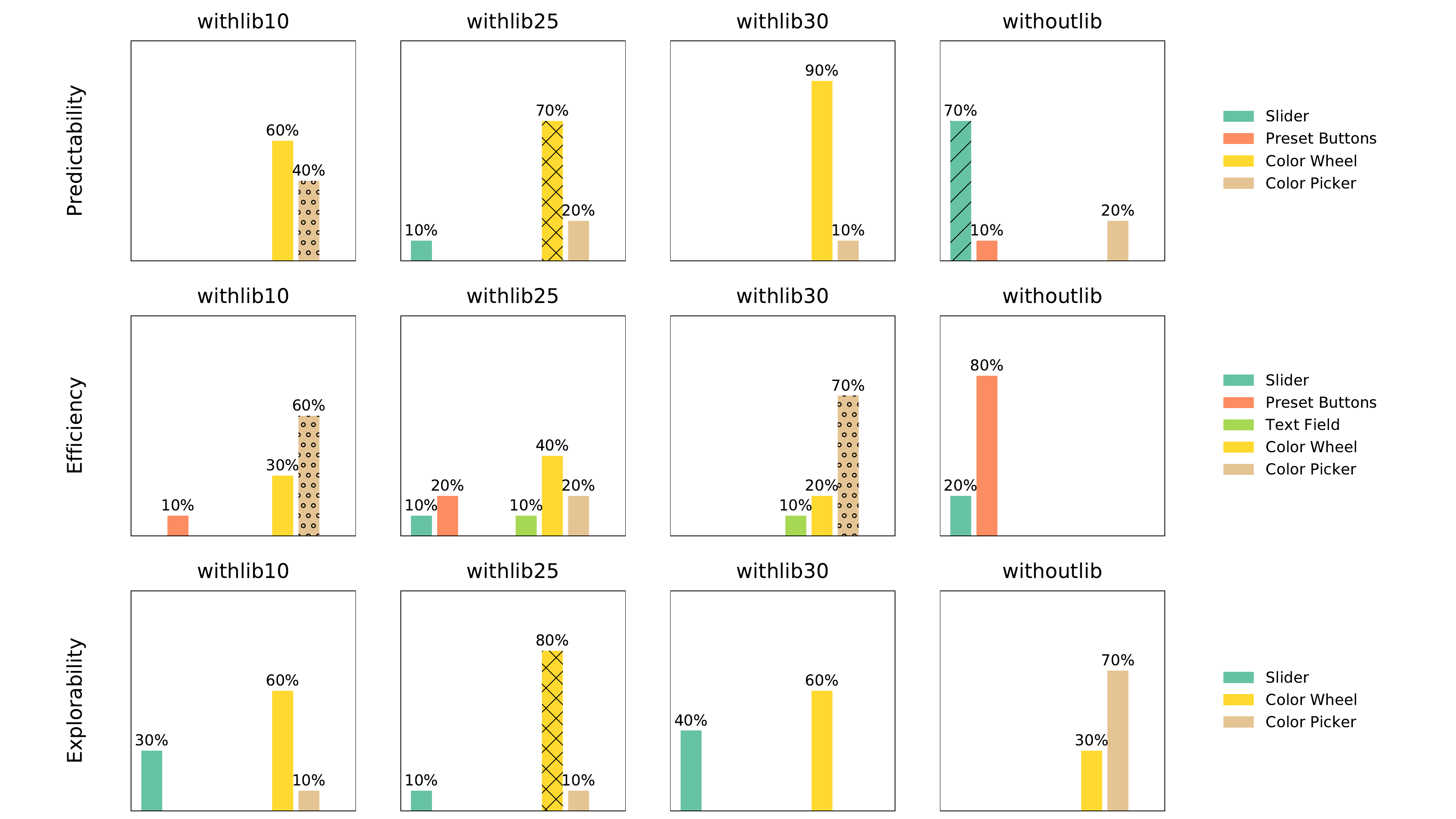}
    \caption{LLM-reasoned UI widgets for \texttt{image\_adjust\_tint} with three sizes of the crowdsourced libraries and without any library. \texttt{Withlib10}, \texttt{withlib25}, and \texttt{withlib30} refer to using the crowdsourced libraries with 10, 25, and 30 user responses for all aspects and tasks, while \texttt{withoutlib} refers to using no library.}
    \Description{This figure consists of three rows, showing the LLM-reasoned widgets for the three preference aspects in each row. Each row has four columns, depicting the reasoned widget frequency as percentages using the four library sizes---withlib10, withlib25, withlib30, withoutlib---in each column as bar charts.}
    \label{fig:reasoning_image_adjust_tint_main}
\end{figure*}

Before we present the study design to address \textbf{RQ3}, we first explain our motivation for studying this question based on an analysis of LLM-generated UI widgets. 
Figure~\ref{fig:reasoning_image_adjust_tint_main} shows an example of LLM-reasoned widgets for task \texttt{image\_adjust\-\_tint} enhanced by three different sizes of the crowdsourced preference library as well as without any library---\texttt{withoutlib}. \texttt{Withlib10}, \texttt{withlib25}, and \texttt{withlib30} stand for the crowdsourced library with 10, 25, and 30 unique responses to each preference aspect. As introduced in Section~\ref{sec:crowdsourcing_result_analysis}, we obtain 30 responses to each aspect in total and use its subsets to construct \texttt{withlib10} and \texttt{withlib\-25}. Due to space limit, we provide the LLM-reasoned widgets with different library sizes for all six tasks in Appendix~\ref{appendix:widgets_of_user_evaluation_tasks}.

As seen in Figure~\ref{fig:reasoning_image_adjust_tint_main} (and figures in Appendix~\ref{appendix:widgets_of_user_evaluation_tasks}), the percentages in the bar charts show the frequency of each type of UI widget reasoned by the LLM over 10 iterations. For instance, in the bar chart for efficiency reasoned by \texttt{withoutlib} (the rightmost chart in the second row of Figure~\ref{fig:reasoning_image_adjust_tint_main}), the preset buttons are reasoned 8 times (80\%), while the slider is reasoned 2 times (20\%). 
Reasoned widgets differing in multiple iterations are caused by the ad-hoc nature of LLMs~\cite{barke2023grounded, vaithilingam2024dynavis}, i.e., they do not always generate all possible widgets in a single pass. This variability is not merely a limitation---it reflects the LLM’s capacity to explore diverse design possibilities through probabilistic reasoning~\cite{ozturkler2022thinksum}, producing a dynamic space of UI designs. 
By contrast, fixed UIs with built-in widget suggestions rely on static, preprogrammed heuristics that prioritize generality. While such systems may use common user preferences for UI suggestion, they limit broader possibilities by enforcing a one-size-fits-all UI design approach, which struggles to accommodate niche workflows or evolving user expectations~\cite{findlater2009design}. Instead, LLM’s dynamic, iterative UI generation surfaces a spectrum of context-aware widgets, transforming variability into a mechanism for adaptive UI suggestions.

Thus, to leverage the design variability enabled by the non-deterministic LLM outputs, we use the reasoning frequency to indicate the recommendation strength for each type of widget by assigning each a score in our user evaluation. For example, the frequencies of 8 for the preset buttons and 2 for the slider are used as scores 8 and 2 for them, as seen in Figure~\ref{fig:ui_image_adjust_tint} Option 2. 
Aggregating widget frequencies across iterations acts as an implicit voting system, where higher-frequency widgets signal stronger consensus of user preference, akin to collaborative filtering in recommender systems~\cite{schafer2007collaborative}. This emergent prioritization reduces the manual effort required to curate or update UI suggestions, as the LLM dynamically reweights suggestions based on reasoned relevance. The scoring of widgets by frequency allows users to quickly identify both common and edge-case tools, mirroring how human designers iteratively refine interfaces through trial and error of multiple possibilities~\cite{nielsen2002iterative} and how recommender systems balance relevance with serendipity when presenting content~\cite{de2015investigation, kotkov2016survey}.

As we observe that the types and frequencies of LLM-reasoned UI widgets vary depending on the size of the library, we are motivated to explore \textbf{RQ3}---the optimal size of the library that maximizes LLM's UI reasoning capabilities. This question is crucial not only for our study but also for broader machine learning research, where determining the optimal dataset size is essential for balancing cost and performance. A well-sized library should be large enough to represent diverse user inputs, but cannot grow indefinitely due to redundancy and cost. Moreover, overly large libraries risk overfitting the model or diluting meaningful patterns in user preference, making it crucial to find a size that is both effective and efficient. To this end, we perform pairwise comparisons of widgets generated with libraries of varying sizes, which are introduced in the following subsection.

\subsubsection{Pairwise Comparisons}
We conduct pairwise comparisons for LLM-generated UI widgets when \texttt{withlib10}, \texttt{withlib25}, \texttt{withlib\-30}, and \texttt{withoutlib} are used. 
Participants need to select their preferences by comprehensively considering the widgets and their frequencies, where a higher frequency indicates a stronger recommendation, and the LLM's reasoning, which explains why each widget is deemed appropriate for the task and preference aspect.
Consequently, each preference aspect in each task results in six pairwise comparisons, as outlined in Table~\ref{tab:pairwise_comparisons}.

\begin{table}[!ht]
    \centering
    \caption{Pairwise comparisons of different crowdsourced library sizes for each preference aspect in each task. Pair orders are randomized in the user evaluation.}
    \Description{This table details the six possible pairwise comparisons of library sizes.}
    \scalebox{0.8} {
    \begin{tabular}{cccc}
    \toprule
        & \multicolumn{3}{c}{\textbf{Library Size Comparison}} \\
        \midrule
        1 & \texttt{withlib10} & \textit{vs.} & \texttt{withlib25} \\
        \midrule
        2 & \texttt{withlib10} & \textit{vs.} & \texttt{withlib30} \\
        \midrule
        3 & \texttt{withlib10} & \textit{vs.} & \texttt{withoutlib} \\
        \midrule
        4 & \texttt{withlib25} & \textit{vs.} & \texttt{withlib30} \\
        \midrule
        5 & \texttt{withlib25} & \textit{vs.} & \texttt{withoutlib} \\
        \midrule
        6 & \texttt{withlib30} & \textit{vs.} & \texttt{withoutlib} \\
    \bottomrule
    \end{tabular}
    }
\label{tab:pairwise_comparisons}
\end{table}

In total, this gives us 6 pairs $\times$ 3 aspects $\times$ 6 tasks = 108 comparisons. To prevent user fatigue, each participant was assigned 18 pairs, which consisted of all six pairs for each aspect across three tasks. Specifically, the user study was divided into two task sets, each containing three tasks. Each participant provided preferences for all three aspects of their assigned task set. The tasks were presented in all six possible permutations to account for order effects, and the sequence of aspects was counterbalanced using the Latin Square method~\cite{colbourn2010crc}. See Table~\ref{tab:user_eval_setup} for the detailed study setup. 

\begin{table*}[!ht]
    \centering
    \caption{User evaluation setup. The six user evaluation tasks are divided into two task sets. Within each set, all six possible permutations of task and aspect orders are employed to control for order effects, and the sequence of pairwise comparisons is randomized.}
     \Description{This table consists of two components: the evaluation setup for task sets 1 and 2. In each component, the sub-table contains three columns: the total order number of tasks, task names, and preference aspects.}
    \scalebox{0.8} {
    \begin{tabular}{cccccc}
    \toprule
        \multicolumn{3}{c}{\textbf{Task Set 1}} & \multicolumn{3}{c}{\textbf{Task Set 2}} \\
        \cmidrule(r){1-3} \cmidrule(l){4-6}
        \textit{Order} & \textit{Task Name} & \textit{Aspect} & \textit{Order} & \textit{Task Name} & \textit{Aspect} \\ 
        \cmidrule(r){1-3} \cmidrule(l){4-6}
        \multirow{4}{*}{6 $\times$} & \texttt{image\_adjust\_exposure} & Predictability & \multirow{4}{*}{6 $\times$} & \texttt{image\_adjust\_tint} & Predictability\\
        \cmidrule(r){2-3} \cmidrule{5-6}
         & \texttt{image\_adjust\_temperature} & Efficiency &  & \texttt{image\_change\_to\_spring} & Efficiency\\
        \cmidrule(r){2-3} \cmidrule{5-6}
         & \texttt{design\_align\_text} & Explorability &  & \texttt{design\_position\_logo} & Explorability\\ 
        \cmidrule(r){1-3} \cmidrule(l){4-6}
        \multirow{4}{*}{6 $\times$} & \texttt{image\_adjust\_exposure} & Efficiency & \multirow{4}{*}{6 $\times$} & \texttt{image\_adjust\_tint} & Efficiency\\
        \cmidrule(r){2-3} \cmidrule{5-6}
         & \texttt{image\_adjust\_temperature} & Explorability &  & \texttt{image\_change\_to\_spring} & Explorability\\
        \cmidrule(r){2-3} \cmidrule{5-6}
         & \texttt{design\_align\_text} & Predictability &  & \texttt{design\_position\_logo} & Predictability\\ 
        \cmidrule(r){1-3} \cmidrule(l){4-6}
        \multirow{4}{*}{6 $\times$} & \texttt{image\_adjust\_exposure} & Explorability & \multirow{4}{*}{6 $\times$} & \texttt{image\_adjust\_tint} & Explorability\\
        \cmidrule(r){2-3} \cmidrule{5-6}
         & \texttt{image\_adjust\_temperature} & Predictability &  & \texttt{image\_change\_to\_spring} & Predictability\\
        \cmidrule(r){2-3} \cmidrule{5-6}
         & \texttt{design\_align\_text} & Efficiency &  & \texttt{design\_position\_logo} & Efficiency\\
    \bottomrule
    \end{tabular}
    }
    \label{tab:user_eval_setup}
\end{table*}

\subsubsection{Data Analysis}
We employ the Chi-squared ($\chi^2$) test~\cite{pearson1900x} to analyze the pairwise comparisons. The $\chi^2$ test evaluates whether observed preference selection frequencies deviate significantly from expected frequencies under the null hypothesis of no association between library size and user preference. We implement this analysis programmatically using SciPy~\cite{scipy} with significance levels denoted by asterisks: $*$ for $p < 0.05$, $**$ for $p < 0.01$, and $***$ for $p < 0.001$.

\subsection{Results}
\subsubsection{Overview of All Tasks}
\label{sec:user_eval_results_per_aspect_all_tasks}
\begin{figure*}[!ht]
    \centering
    \begin{subfigure}[b]{0.75\textwidth}
        \includegraphics[width=\textwidth]{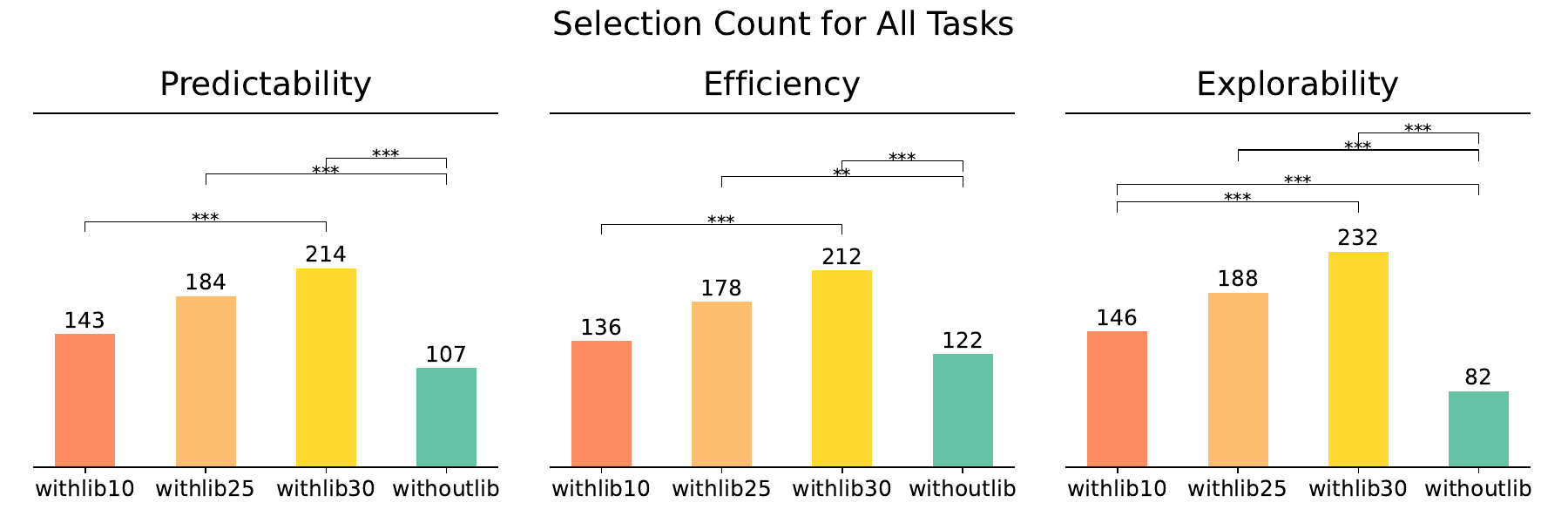}
        \caption{Total selection counts.}
        \label{fig:user_eval_per_aspect}
    \end{subfigure}
    \par\vspace{0.3cm}
    \begin{subfigure}[b]{0.75\textwidth}
        \includegraphics[width=\textwidth]{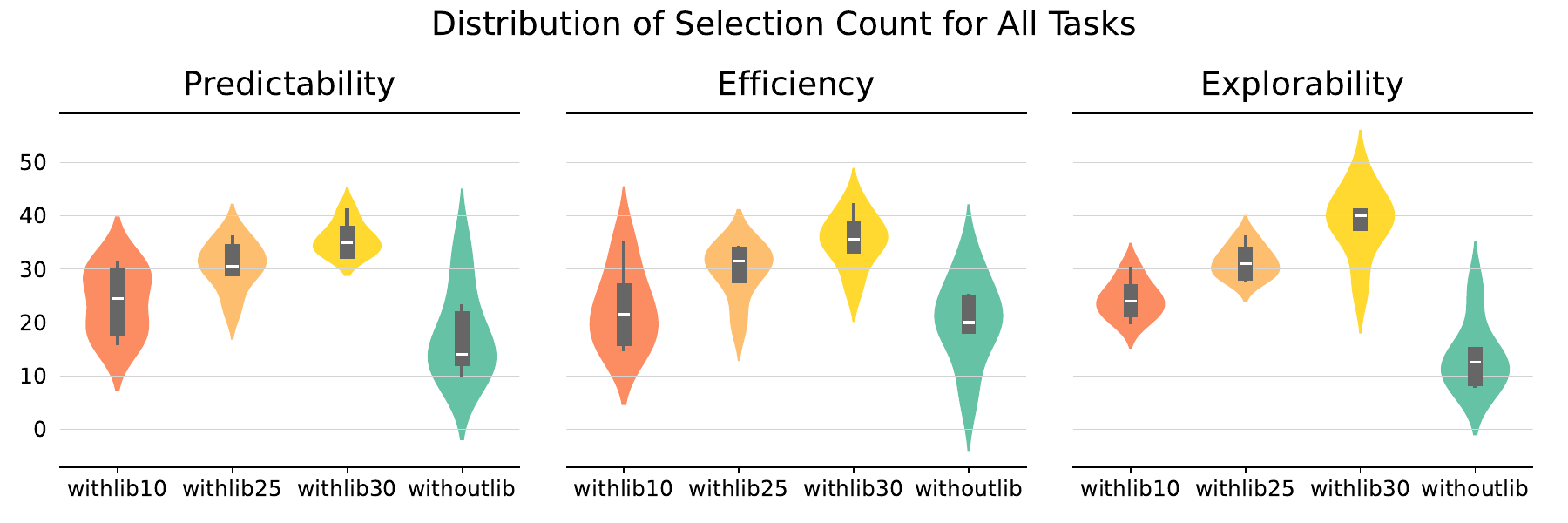}
        \caption{Distribution of total selection counts. The box plot elements show the median, quartiles (25th and 75th percentiles), and the data range.}
        \label{fig:user_eval_per_aspect_violin}
    \end{subfigure}
    \caption{Number and distribution of user preference selection counts of three preference aspects across all tasks.}
    \Description{This figure contains two sub-figures: (a) shows the total selection counts as bar charts; (b) shows the distribution of total selection counts as violin plots. Both sub-figures consist of three columns that present the results of the three preference aspects.}
\end{figure*}

We present the user evaluation results of each preference aspect across all tasks in Figures~\ref{fig:user_eval_per_aspect} and \ref{fig:user_eval_per_aspect_violin}. For all three aspects, users preferred the LLM-reasoned widgets with libraries over those without any library. When the libraries were incorporated, \texttt{withlib30} was consistently more preferred than \texttt{withlib10} and \texttt{withlib25}.

\subsubsection{Inspect Each Task}
\label{sec:user_eval_results_per_task}
We further present the results of each evaluated task. According to the results, user preferences show varying trends depending on the task.

\begin{figure*}[!ht]
    \centering
    \begin{subfigure}[b]{0.75\textwidth}
        \includegraphics[width=\textwidth]{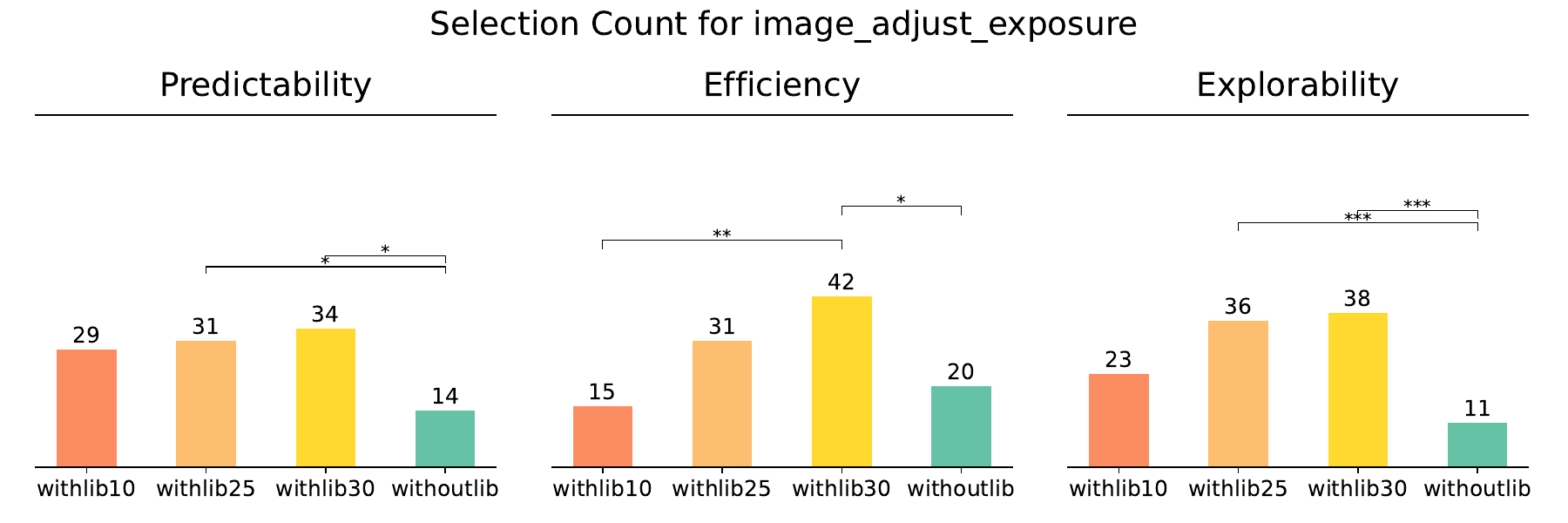}
    \end{subfigure}
    \par\vspace{0.3cm}
    \begin{subfigure}[b]{0.75\textwidth}
        \includegraphics[width=\textwidth]{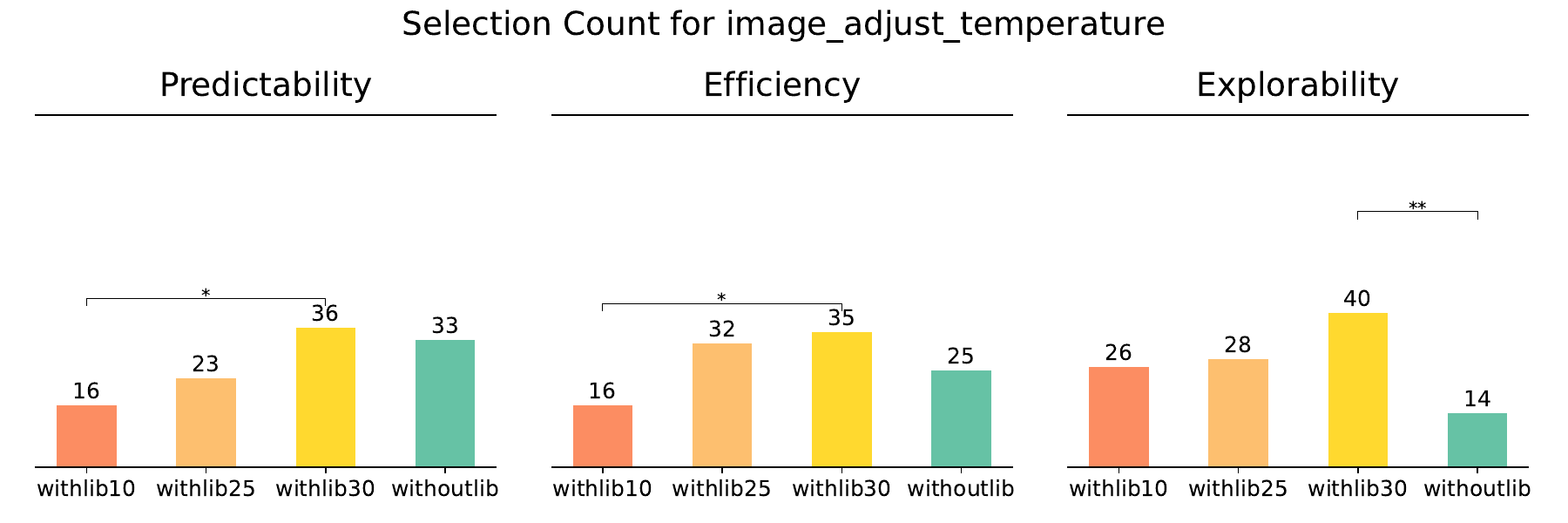}
    \end{subfigure}
    \par\vspace{0.3cm}
    \begin{subfigure}[b]{0.75\textwidth}
        \includegraphics[width=\textwidth]{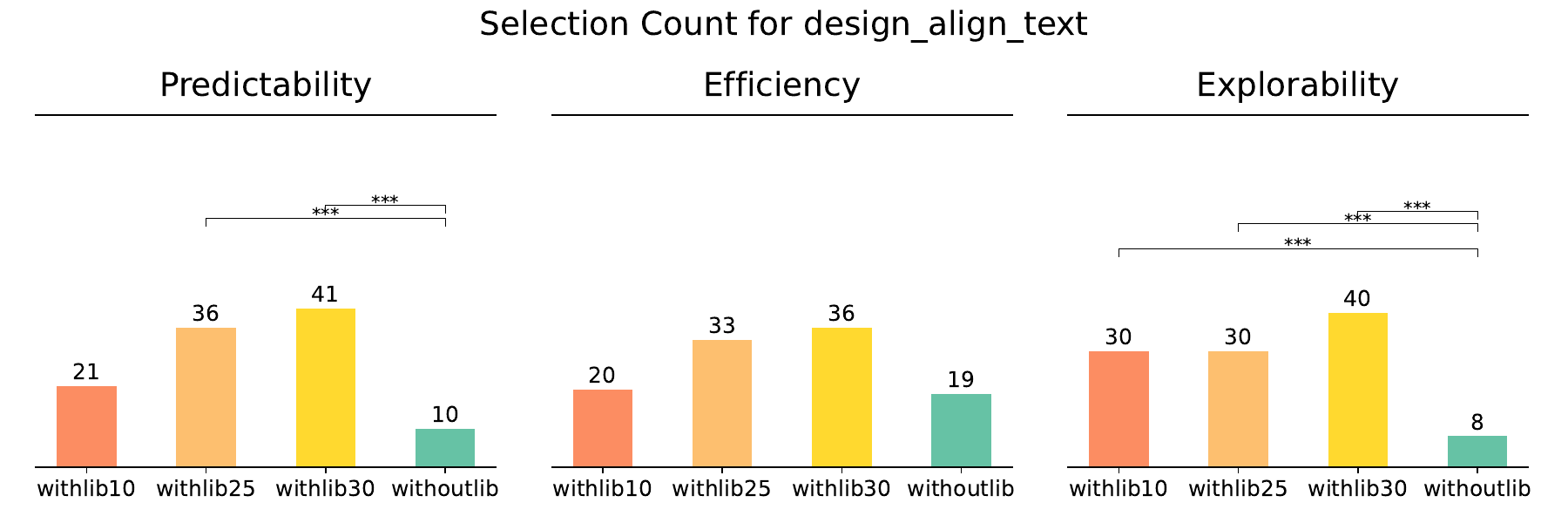}
    \end{subfigure}
    \caption{User preference selection counts of tasks in task set 1.}
    \Description{This figure contains three sub-figures in three rows, showing the total selection counts as bar charts for tasks in task set 1. All sub-figures consist of three columns that present the results of the three preference aspects.}
    \label{fig:user_eval_per_task_set1}
\end{figure*}

\begin{figure*}[!ht]
    \centering
    \begin{subfigure}[b]{0.75\textwidth}
        \includegraphics[width=\textwidth]{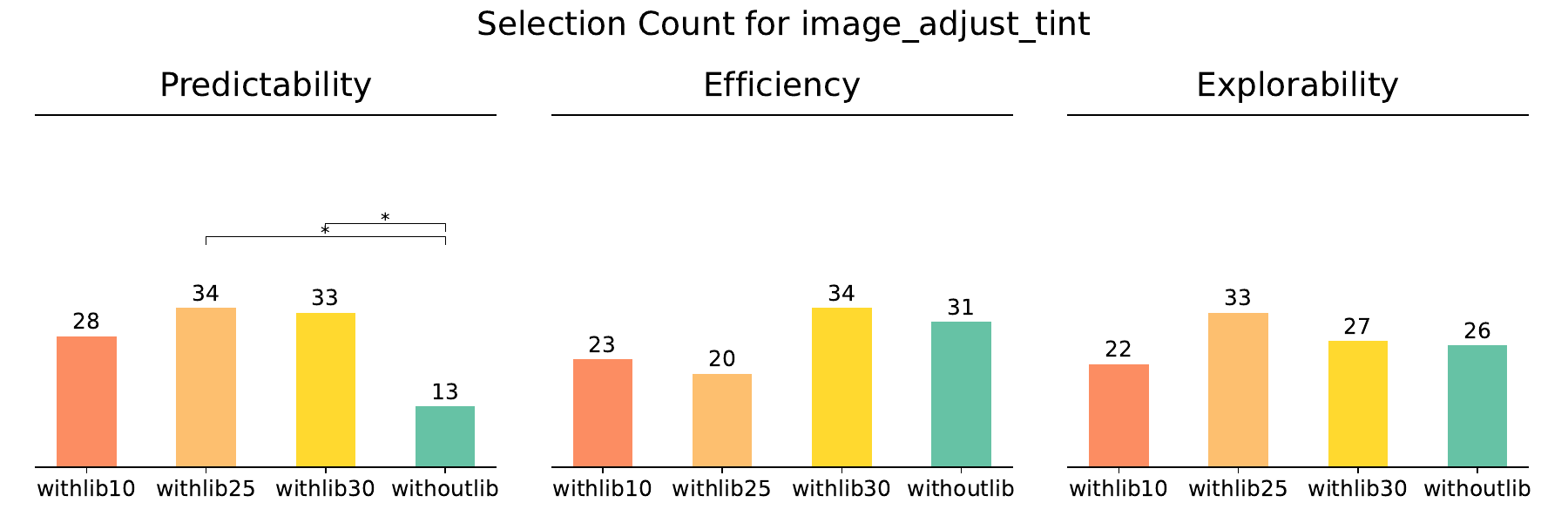}
    \end{subfigure}
    \par\vspace{0.3cm}
    \begin{subfigure}[b]{0.75\textwidth}
        \includegraphics[width=\textwidth]{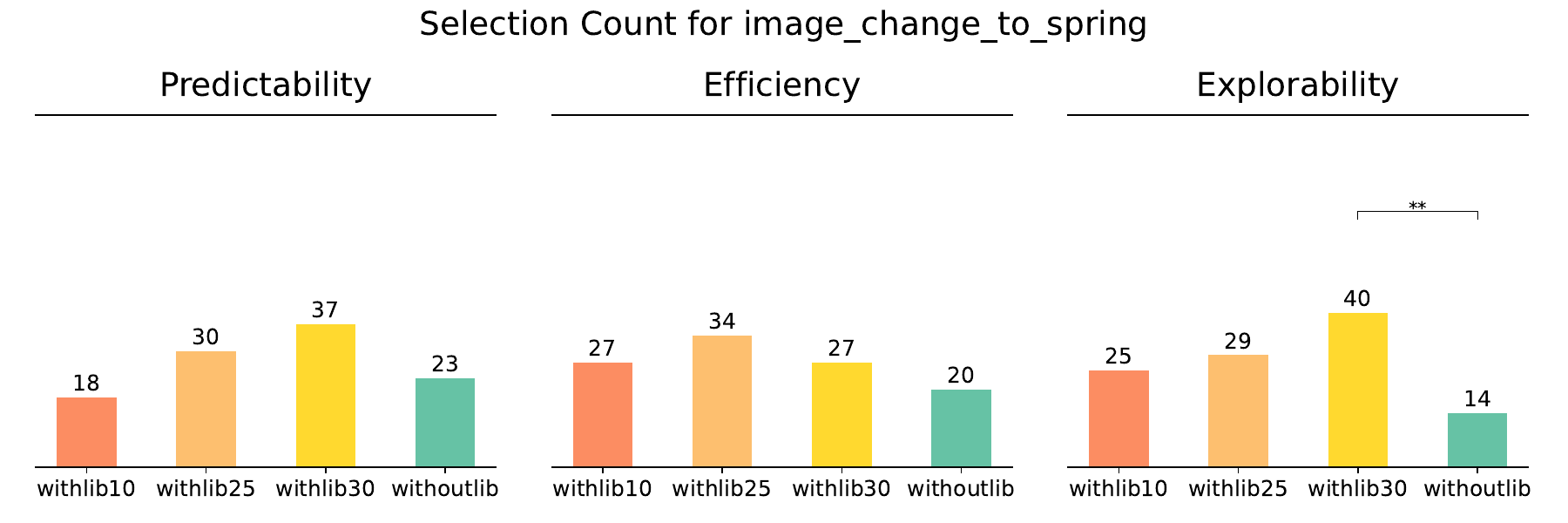}
    \end{subfigure}
    \par\vspace{0.3cm}
    \begin{subfigure}[b]{0.75\textwidth}
        \includegraphics[width=\textwidth]{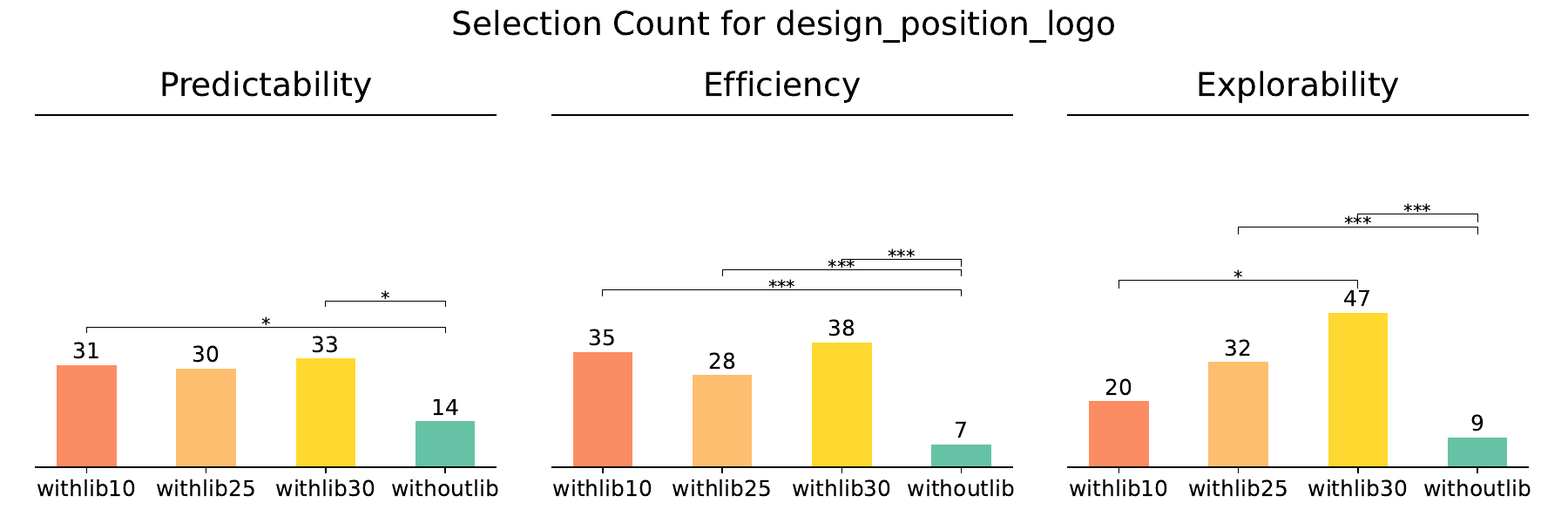}
    \end{subfigure}
    \caption{User preference selection counts of tasks in task set 2.}
    \Description{This figure contains three sub-figures in three rows, showing the total selection counts as bar charts for tasks in task set 2. All sub-figures consist of three columns that present the results of the three preference aspects.}
    \label{fig:user_eval_per_task_set2}
\end{figure*}

\textbf{Figure~\ref{fig:user_eval_per_task_set1}---task set 1}: 

\begin{itemize}[left=0pt]
    \item \texttt{image\_adjust\_exposure} (see Figure~\ref{fig:reasoning_image_adjust_exposure} for LLM-reasoned widgets): Widgets reasoned with libraries were more preferred than those without a library for both predictability and explorability, with \texttt{withlib25} and \texttt{withlib30} being nearly equally favored. For efficiency, \texttt{withlib25} and \texttt{withlib30} were preferred over \texttt{withoutlib}, and \texttt{withoutlib} was favored over \texttt{withlib10}. 
    
    \item \texttt{image\_adjust\_temperature} (see Figure~\ref{fig:reasoning_image_adjust_temperature} for LLM-reason\-ed widgets): \texttt{Withlib30} was the most favored and almost equally favored as \texttt{withoutlib} for predictability and \texttt{withlib25} for efficiency. In addition, widgets reasoned with libraries were consistently favored for explorability. 
    
    \item \texttt{design\_align\_text} (see Figure~\ref{fig:reasoning_design_align_text} for LLM-reasoned widgets): Widgets reasoned with libraries were consistently more favored for all three aspects.
\end{itemize}

\textbf{Figure~\ref{fig:user_eval_per_task_set2}---task set 2}: 
\begin{itemize}[left=0pt]
    \item \texttt{image\_adjust\_tint} (see Figure~\ref{fig:reasoning_image_adjust_tint_main} for LLM-reasoned widgets): Widgets reasoned with libraries were more favored for predictability compared to those without a library. For efficiency, \texttt{withlib30} and \texttt{withoutlib} were almost equally preferred. Regarding explorability, \texttt{withlib25} was the most favored.
    
    \item \texttt{image\_change\_to\_spring} (see Figure~\ref{fig:reasoning_image_change_to_spring} for LLM-reasoned widgets): \texttt{With\-lib30} was the most preferred for predictability and explorability. In terms of efficiency, \texttt{withlib25} was preferred over the other libraries. 
    
    \item \texttt{design\_position\_logo} (see Figure~\ref{fig:reasoning_design_position_logo} for LLM-reaso\-ned widgets): Widgets reasoned with libraries were significantly more favored across all three aspects. 
\end{itemize}

\subsection{Findings}
\label{sec:user_eval_results_reflection}
\subsubsection{Generalization to Nonoverlapping Tasks and Broader Domains [RQ1]}
The results reveal that the user preferences collected from crowdsourcing can effectively generalize to tasks beyond the original crowdsourcing tasks by enhancing LLM reasoning. Across all user evaluation tasks, including the tasks not in the image editing domain, widgets reasoned with the library are largely favored over those reasoned without a library. This highlights the adaptability and relevance of user preferences captured through crowdsourcing, suggesting that these insights can serve as a foundation to inform widget reasoning across a range of related tasks. 

\textbf{Takeaway}: Crowdsourced user preferences effectively enhance LLM widget reasoning, not only being favored across image editing tasks but also generalizing to inform widget reasoning for relevant tasks beyond image editing.

\subsubsection{Preference Alignment of Generated UI Enhanced by Crowdsourced Library [RQ2]}
The results demonstrate that incorporating the crowdsourced library enhances the LLM's ability to reason widgets that align with user preferences. 
For predictability, users' preferences captured from the crowdsourcing lean towards preset buttons for predefined value selection, color wheels, and color pickers for color adjustments, and click on image for position changes (recall Section~\ref{sec:crowdsourcing_result_analysis}). These preferences have directly influenced the widgets reasoned for the user evaluation tasks, where widgets reasoned with the library adhered closely to these favored widget types. In contrast, when the library is not used, the LLM reasons sliders for most tasks regardless of the task categories. 
For efficiency, similar patterns emerge---preset buttons, color wheels, color pickers, and click on image are repeatedly preferred for their ease and speed in achieving task goals. The LLM, when equipped with the crowdsourced library, can reason toward these efficient widgets, whereas, without the library, the reasoning leans more heavily on sliders and preset buttons for most tasks without too much differentiation for task categories. 
Regarding explorability, the user preference trends show that sliders are favored for continuous value adjustments, color wheels are preferred for color changes, and click on image is liked for position adjustments. Without the library, however, the LLM tends to reason sliders to allow exploration for continuous value adjustments, color pickers for color changes, and sliders and text fields for position adjustments. The incorporation of the crowdsourced library enables the LLM to capture the need for exploration in tasks like color and position adjustment by aligning the widget choices with user-preferred ones.

\textbf{Takeaway}: Integrating the crowdsourced library enables the LLM to reason widgets that more accurately align with user preferences for predictability, efficiency, and explorability, while its absence leads to more generic and less tailored reasoning.

\subsubsection{Effect of Library Sizes on UI Generation [RQ3]}
Both the total selection counts and their distributions, shown in Figures~\ref{fig:user_eval_per_aspect} and \ref{fig:user_eval_per_aspect_violin}, highlight \texttt{withlib30} as the most reliable for all three aspects due to the richness of its crowdsourced data. This consistency of user preferences underscores the value of extensive, diverse data in reasoning user-aligned widgets.
On the other hand, \texttt{withoutlib} shows significant variability in user preferences, suggesting that when no library is used, LLM reasoning lacks the consistency that is otherwise afforded by crowdsourced data. 
However, examining each task individually reveals some variation in preferred libraries for different preference aspects. This indicates that while larger libraries lead to consistent preference-aligned widget reasoning in general, tasks like \texttt{image\_adjust\_temperature} and \texttt{image\_adjust\_tint} show that \texttt{withoutlib} is almost equally favored as \texttt{withlib30} regarding certain aspects and may need additional evaluation.

\textbf{Takeaway}: Extensive crowdsourced data makes it the most reliable and consistent for widget reasoning across tasks, while libraries with fewer data lead to greater variability in user preferences.

\section{Discussion}
\label{sec:discussion}
In this section, we reflect on our findings and discuss their implications for crowdsourcing-enhanced human-centered UI generation. 

\subsection{Improving Task Generalizability with Curated Crowdsourcing Setup}
\textbf{Importance of task curation to collect helpful user data that can boost LLM-based UI generation}:
The user evaluation findings indicate that the crowdsourced user preference library improved the LLM's reasoning capabilities, extending the library's effectiveness beyond its initial intended task scope and domain. 
Such generalizability highlights the user data's potential to serve as a foundational element to guide the creation of versatile, human-centered UIs across diverse domains through careful task curation. This can contribute to a significant reduction in the necessity for task-specific UI customizations. For instance, if we collect user-preferred UIs to adjust image tones, this insight can be helpful in indicating user preferences for a range of similar tasks involving continuous color adjustment. In our implementation, the effectiveness of incorporating tasks belonging to the three categories involving diverse data manipulation requirements (recall Section~\ref{sec:crowdsourcing_tasks}) has been demonstrated through our evaluation. This implies that through well-curated task design, UI designers can collaborate with LLMs that are equipped with user data to prototype designs to meet users' expectations with reduced additional user testing. 

\textbf{Extending the breadth and depth of task categories}: 
To further enhance the user data's generalizability, future efforts can strategically encompass a broader spectrum of task categories. In addition to this work's focus on image parameter adjustment, other potential task categories include audio, video, data visualization editing, and others introduced in Section~\ref{sec:relatedwork}. These domains often involve distinct interaction styles. For instance, video editing might require precise temporal controls like scrubbers and frame-level sliders, while data visualizations might prefer touch-based navigation on tablets versus hover-driven interactions on desktops. Because each domain brings unique data manipulation, their UI designs may not be easily generalizable across domains. As such, multi-domain user data collection is essential to ensure UI recommendations remain relevant and effective for diverse task contexts.

On the other hand, future research can also aim for a deeper exploration of specific task categories, focusing on more complex and intricate forms of data manipulation. For instance, for image editing, we can extend the scope from single-value to multi-dimensional or compound parameter manipulations. This is a common need as image editing often consists of a combination of multiple parameters, e.g., if a user wants to correct a low-exposure image, they may increase both image saturation and brightness. Thus, a dataset containing user data on complex or open-ended tasks may help inform the LLM-based UI generation framework to create a widget that combines multi-parameter adjustment or to generate widget combinations. Investigating extended task complexities can help uncover the dependence of specific parameters and the relevance of high-level task requirements, thus refining the granularity of crowdsourced data to bolster the applicability of LLM-generated UIs in addressing complicated task contexts. 

\subsection{Designing User-Focused UIs with Insights Collected from Target Audience}
\textbf{Tailoring UIs to target users favored over traditional fixed UI designs}:
Our crowdsourcing results highlight variations in user preferences for UI widgets, underscoring the need for a user-centered approach to widget selection. This contrasts with traditional ways of UI design by most software and applications, i.e., UIs are fixed and consist of designers' pre-selected UI elements. For instance, Photoshop offers a slider and text field for adjusting image lightness, saturation, and hue. However, our crowdsourced data reveals various user-preferred widgets that go beyond these given options. This divergence exposes a gap that static, designer-selected widgets sometimes fail to meet diverse user needs. To this end, our framework utilizes LLM-based dynamic UI design, which generates UIs aligned with user preferences and task contexts. This framework facilitates the creation of user-centered interfaces with direct end-user input versus the traditional designer-driven methods.

\textbf{Collecting specific, targeted user data to guide tailored UI design}:
Our user-centered framework can further evolve to encompass a broader range of preference dimensions and serve more diverse user populations. While this work uses three preference aspects to demonstrate the framework's efficacy, future work can explore additional dimensions by using established UI design principles~\cite{mckay2013ui, lidwell2010universal, chen2021should, shneiderman2000creating, norman1983design}. Among them, we take accessibility as an example to show how other dimensions can be incorporated. Accessible UI design is a crucial dimension and is addressable through crowdsourcing preferences from users with disabilities. The collected data would enable LLMs to reason accessibility considerations directly based on the target user's data. For instance, users with visual impairments might prefer high-contrast color schemes, screen reader-friendly labels, or voice-based navigation. Those with motor impairments may favor interfaces with larger clickable targets, simplified gesture controls, or customizable input methods based on varied motor abilities~\cite {gajos2004supple, gajos2007automatically}. These accessibility preferences, once incorporated into the user library, can be used to guide LLM reasoning during UI generation. In turn, the LLM can recommend design adaptations tailored to specific user groups, such as suggesting larger widget sizes or reducing the need for precise cursor movements, and provide explanations grounded in the collected data. This not only supports more inclusive UI outcomes for end users but also makes the rationale behind design recommendations transparent for designers' reference. 

\subsection{Building Adaptive UIs with Scalable User Data}
\textbf{Multi-scale user data to guide LLM reasoning on UI designs}:
Based on the comparisons of libraries of varying sizes, our results reveal that larger libraries generally yield consistent preference-aligned widget reasoning across diverse tasks. On the other hand, smaller libraries stand out for their cost-effectiveness and adaptability to guide UI design. Our study demonstrates that a library with as few as 10 user responses can effectively capture insights into user preferences. In specific tasks, such as adjusting image exposure, shifting colors to spring tones, and positioning logos, small-scale libraries perform comparably with their larger counterparts for certain preference aspects. These libraries are quick to compile, offering users and designers fast UI design recommendations without the need for extensive data collection and user testing. This highlights how LLMs can support UI design with cost-efficient human input for user-centered design. 

\textbf{Low-cost, small-scale user data for UI personalization and evolution}:
The adaptability of small-scale libraries can enable deeper personalization and long-term UI evolution. By incorporating a few user-specific samples, the small data can be used to reflect individual preferences and tailor workflows that align with unique needs. Such UI personalization and evolution can be especially beneficial for beginner users who want to use professional, feature-rich software without feeling overwhelmed at the start. As users’ skills and expertise develop, their personal data can evolve and dynamically adjust the UIs to be more advanced over time. For instance, we can use our framework to initially generate simple, beginner-friendly UIs, then progressively adapt to offer sophisticated UIs with advanced tools as the user’s proficiency grows, guided by an updated personal library covering user traces. This evolution monitors user growth and fosters a gradual and personalized learning curve, allowing users to engage with increasingly complex tools at their own pace. 
With a similar approach, creators can also use UI personalization and evolution to tailor UIs to individual workflows. Although existing software allows users to show/hide tools for UI personalization~\cite{photoshop, blender}, this often does not allow transformative UI redesign like regrouping tools into a user-created panel. Instead, using AI to generate UIs can programmatically create tools with personalized, evolving functionality that is tailored to creators' own needs. This dynamic approach can be beneficial to enhance productivity as users are not required to learn and adapt to given UIs, but instead, UIs are tailored to their workflows. 

\subsection{Exploring UI Design in Human-Centered Agentic AI}
\textbf{Indispensable role of UIs in agentic AI}:
As agentic AI increasingly automates complex tasks, e.g., building agents with Magentic-One~\cite{fourney2024magenticone} to write and execute code, we envision agentic AI systems to help delegate tasks that can significantly foster human productivity. In the meantime, alongside the pursuit of fully autonomous agentic AI systems, designing human-AI collaboration systems remains a critical problem in balancing efficiency with human agency. While agentic systems can autonomously execute actions on behalf of humans, their effectiveness hinges on aligning with user expectations and contextual needs. This challenge is particularly pronounced in domains requiring creativity and exploration. Taking image editing as an example, a task involving image parameter adjustment can yield numerous solutions that different creators come up with. Thus, agentic AI that automates the parameter adjustment can easily fall short of satisfying such subjective and aesthetic needs and lead to numerous iterations of prompt refinement. Instead, UIs with direct controls that bridge the communication between humans and agentic AI become essential. Thus, we argue that the user interface remains indispensable in agentic AI systems where meaningful human engagement is desired.

\textbf{Dataset injection to LLMs as a scalable solution for UI generation to support human-centered agentic AI}:
To allow interface interaction in agentic AI, LLM-generated UIs present a dynamic solution well-suited to the automation and adaptability required by LLM-based AI agents. However, using LLMs trained on generic data alone risks homogenizing design choices to generic solutions unless explicitly guided by target user-driven insights, as shown in this work. This highlights the need for UI generation frameworks to prioritize user needs. Among existing LLM post-training techniques for customizing LLMs to specific tasks and user preferences \cite{kumar2025llm, tie2025survey}, our framework contributes an approach of embedding a crowdsourced user dataset to guide LLMs to produce user-centric outputs through structured reasoning enhanced by this dataset. 
More interestingly, this framework can further benefit from agentic AI systems to deliver UI designs leveraging agent collaboration. Imagine that we can have multiple LLMs working simultaneously to generate UI designs, each equipped with user data covering different user profiles and design principles. A multi-agent system can be built to analyze these LLMs' outputs, coordinate conflicts, and produce UI design suggestions with multiple perspectives considered. This agentic collaboration can enhance coverage and diversity in UI design, as well as support dynamic dialogue between agents to enable emergent design creativity.

\section{Limitations and Future Work}
\label{sec:limitations}

\textbf{New interaction modalities beyond traditional GUIs}: Although our work primarily focused on 2D UI widget generation, an emerging area is applying the framework to novel interaction modalities beyond 2D GUIs for preference alignment. For instance, as voice-controlled interfaces, AR/VR environments, and wearable device UIs become increasingly popular, it would be interesting to collect user preferences for those contexts. Future work can crowdsource user data regarding the voice command structure for a given function based on what most children find intuitive, gestures, or holographic controls that most senior people feel natural in AR, and gesture-based shortcuts or low-interruption interfaces for smartwatches suitable for most people in office settings. This direction can extend the crowdsourcing-informed interaction design as UIs diversify with technology advancements and user contexts.

\textbf{Whole-interface generation and layout}: This work focuses on individual UI widget generation, and future research can expand to the complete interface layout or workflow generation by investigating how a crowdsourced dataset scales to decisions about arranging multiple widgets, navigation structures, or multi-step task flows. For example, given a complex task (e.g., ``design a photo editing toolbar''), what interface layouts do users prefer, and how can an LLM assemble those using collected user insights? Crowdsourcing could be used to gather user-preferred layouts or sequences in addition to widget types. This can extend LLM-based UI generation to entire UIs that are human-centered at a macro level and also opens questions about representing and learning from more complex user data, e.g., sequences of interactions and grouping of controls, and ensuring coherence in generated designs.

\textbf{Integration with design guidelines and experts}: Future work can combine the current crowdsourced general user data with established HCI design principles or input from professional UI designers. Research might explore hybrid systems where the LLM is guided by both the crowd-derived library and rule-based heuristics or where crowd preferences are augmented with expert critiques. This addresses whether certain nuanced design decisions (e.g., accessibility considerations or aesthetic consistency) might require more than just crowd preferences. By involving experts in the loop, e.g., crowdsourcing from designers or using expert-verified crowd data, the resulting UI generation could achieve a balance between general user appeal and professional best practices. 

\textbf{Cross-cultural user data collection}: The current work crowdsourced user preferences primarily from English-speaking participants. This can lead to overlooked preferences and cultural nuances that emerge in broader global contexts. For UI design, user preferences for color schemes, layout conventions, or even interaction metaphors may differ substantially across linguistic and cultural divides. Future data collection could employ multilingual crowdsourcing platforms to enhance cross-cultural inclusiveness and adapt tasks to reflect culturally relevant use cases. This would allow LLMs to account for cross-cultural insights to generate UIs that reflect the diversity of global user populations, thus promoting equitable and context-aware UI design on a global scale.

\textbf{User persona-aware UI design}: Beyond user preferences, future research can incorporate broader user persona data, such as demographic information (e.g., age, gender), domain expertise (e.g., novice vs. expert users), device usage habits, cognitive styles, and other aspects, to represent essential aspects of target users. By leveraging these diverse persona attributes, the framework can be extended to process persona data and further generate UI designs that are not only preference-aligned but also contextually appropriate for different user types. 

\textbf{Technical advancements of using LLMs for UI generation}: This work implements a system prototype of the proposed framework and opens up opportunities for technical improvements. First, using LLMs for code generation introduces limitations, as minor bugs may occur and require manual fixes. Recent research has developed LLM-based agents to fix bugs in code automatically~\cite{yang2024swe, zhang2024autocoderover, kang2025explainable}. This offers a promising solution to robustly generate UIs in future iterations. 
Next, we use GPT-4o as the LLM for UI generation, and future research could explore more advanced LLMs for improved reasoning capabilities, open-source LLMs for lower cost and easier replicability, or fine-tuned LLMs for specific task domains. This would also lead to an in-depth comparison of different LLMs' performance for UI generation.

\section{Conclusion}
\label{sec:conclusion}
In this paper, we introduced the \crowdgenui{} framework for LLM-based widget generation enhanced by a crowdsourced user preference library. Through a user evaluation of the system prototype implemented for this framework, we demonstrated its generalizability to extend widget generation across diverse and related tasks, effectiveness to align generated widgets with user preferences, and capability to use large libraries for consistent preference-aligned widget reasoning and small libraries for cost-effective preference capture. We further discussed the user study results to explore how the framework can be extended to accommodate broader task domains, various user data, and scalable data sizes to support UI generation for emerging task contexts, diverse user needs, and adaptive, user-focused workflows. Based on the discussion, we suggested multiple future research directions regarding LLM-based UI generation and design. We envision our framework to benefit both end users and designers by simplifying the development of human-centered UIs that enhance task efficiency and user satisfaction.



\bibliographystyle{ACM-Reference-Format}
\bibliography{bibliography}

\appendix

\newpage

\section{LLM-Reasoned UI Widgets for User Evaluation Tasks}
\label{appendix:widgets_of_user_evaluation_tasks}
\begin{figure*}[!ht]
    \centering
    \includegraphics[width=0.85\textwidth]{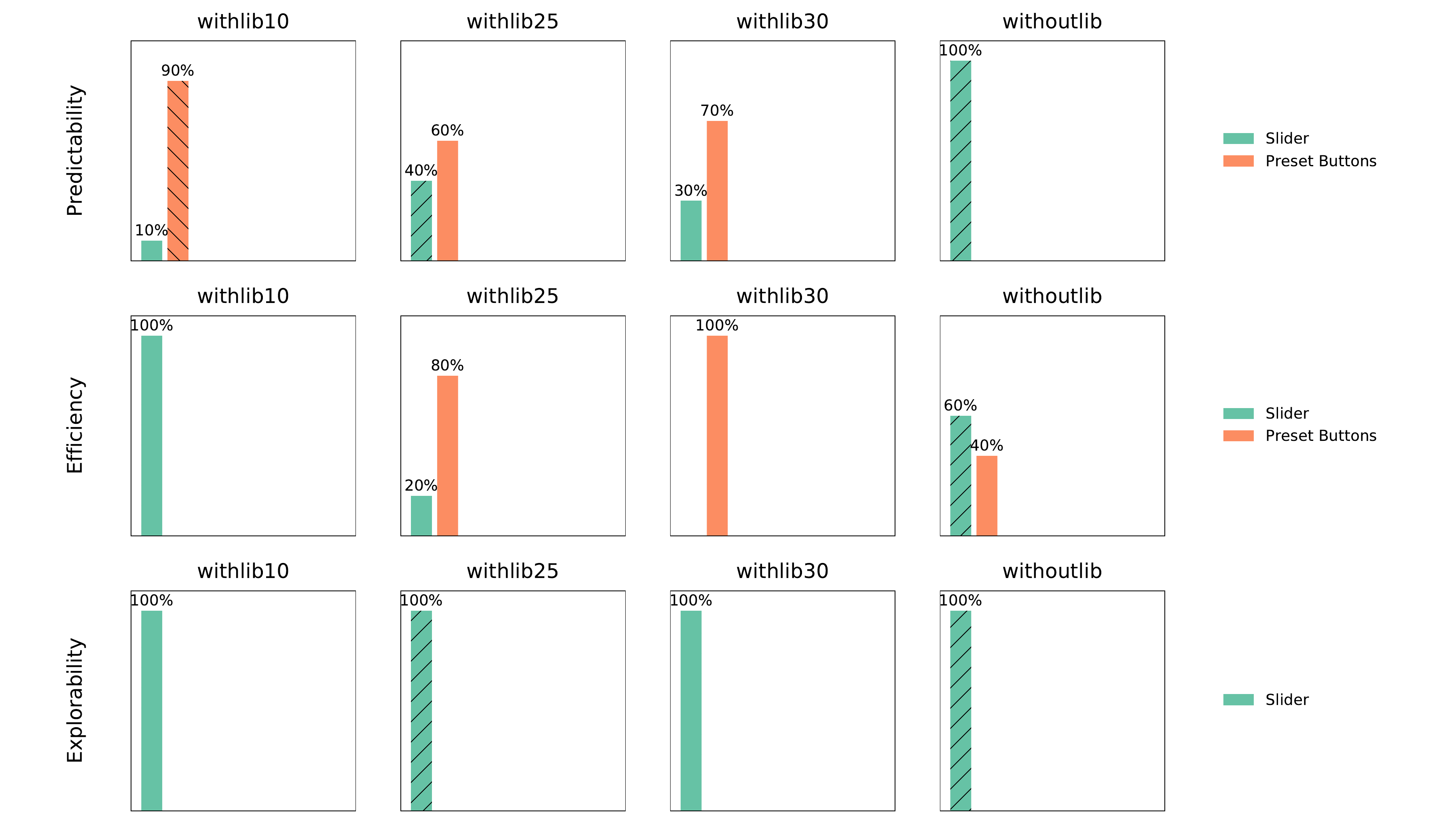}
    \caption{LLM-reasoned UI widgets for \texttt{image\_adjust\_exposure} with three sizes of the crowdsourced libraries and without any library. \texttt{Withlib10}, \texttt{withlib25}, and \texttt{withlib30} refer to using the crowdsourced libraries with 10, 25, and 30 user responses for all aspects and tasks, while \texttt{withoutlib} refers to using no library.}
    \label{fig:reasoning_image_adjust_exposure}
\end{figure*}

\begin{figure*}[!ht]
    \centering
    \includegraphics[width=0.85\textwidth]{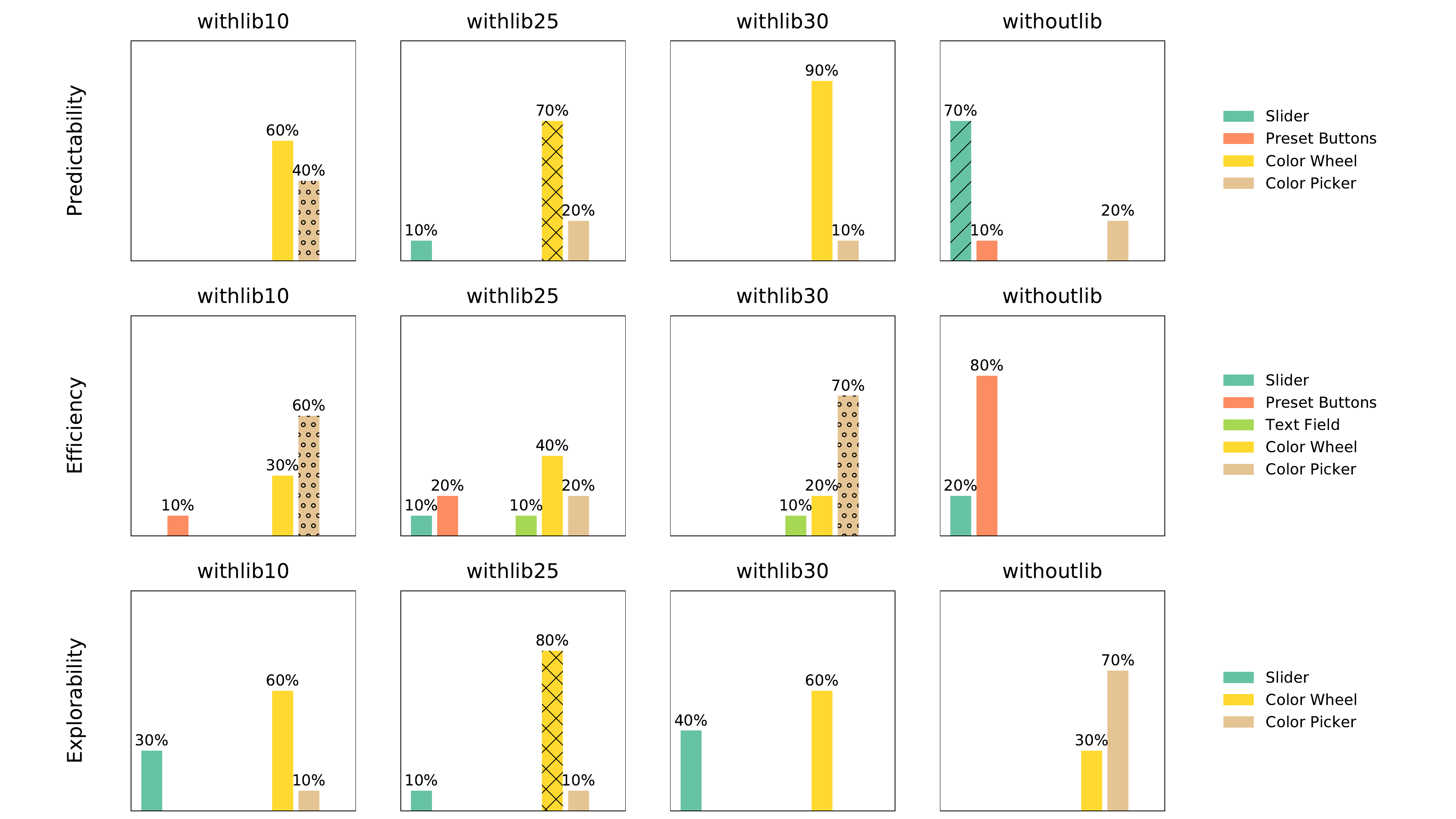}
    \caption{LLM-reasoned UI widgets for \texttt{image\_adjust\_tint} with three sizes of the crowdsourced libraries and without any library. \texttt{Withlib10}, \texttt{withlib25}, and \texttt{withlib30} refer to using the crowdsourced libraries with 10, 25, and 30 user responses for all aspects and tasks, while \texttt{withoutlib} refers to using no library.}
    \label{fig:reasoning_image_adjust_tint}
\end{figure*}

\begin{figure*}[!ht]
    \centering
    \includegraphics[width=0.85\textwidth]{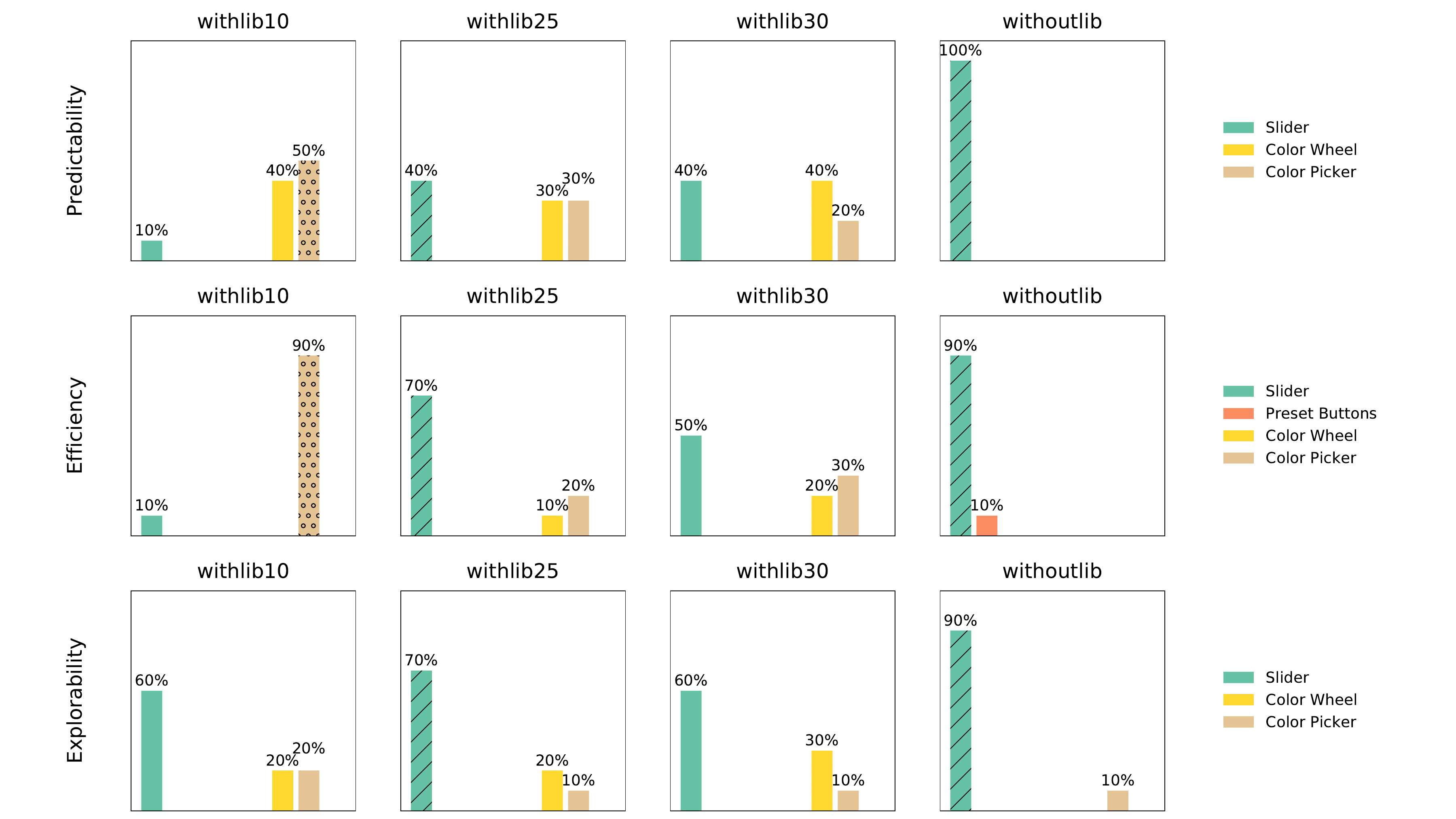}
    \caption{LLM-reasoned UI widgets for \texttt{image\_adjust\_temperature} with three sizes of the crowdsourced libraries and without any library. \texttt{Withlib10}, \texttt{withlib25}, and \texttt{withlib30} refer to using the crowdsourced libraries with 10, 25, and 30 user responses for all aspects and tasks, while \texttt{withoutlib} refers to using no library.}
    \label{fig:reasoning_image_adjust_temperature}
\end{figure*}

\begin{figure*}[!ht]
    \centering
    \includegraphics[width=0.85\textwidth]{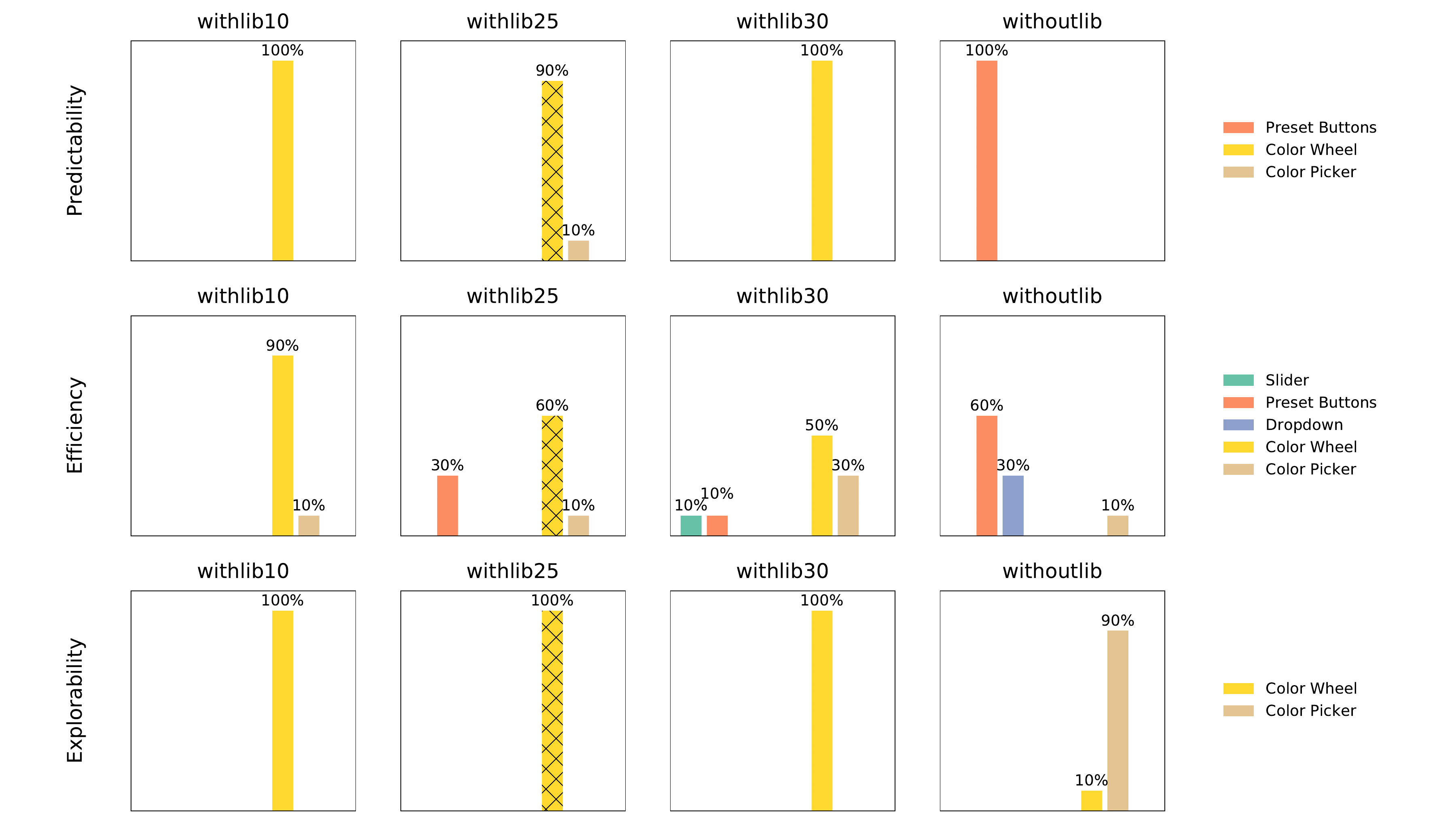}
    \caption{LLM-reasoned UI widgets for \texttt{image\_change\_to\_spring} with three sizes of the crowdsourced libraries and without any library. \texttt{Withlib10}, \texttt{withlib25}, and \texttt{withlib30} refer to using the crowdsourced libraries with 10, 25, and 30 user responses for all aspects and tasks, while \texttt{withoutlib} refers to using no library.}
    \label{fig:reasoning_image_change_to_spring}
\end{figure*}

\begin{figure*}[!ht]
    \centering
    \includegraphics[width=0.85\textwidth]{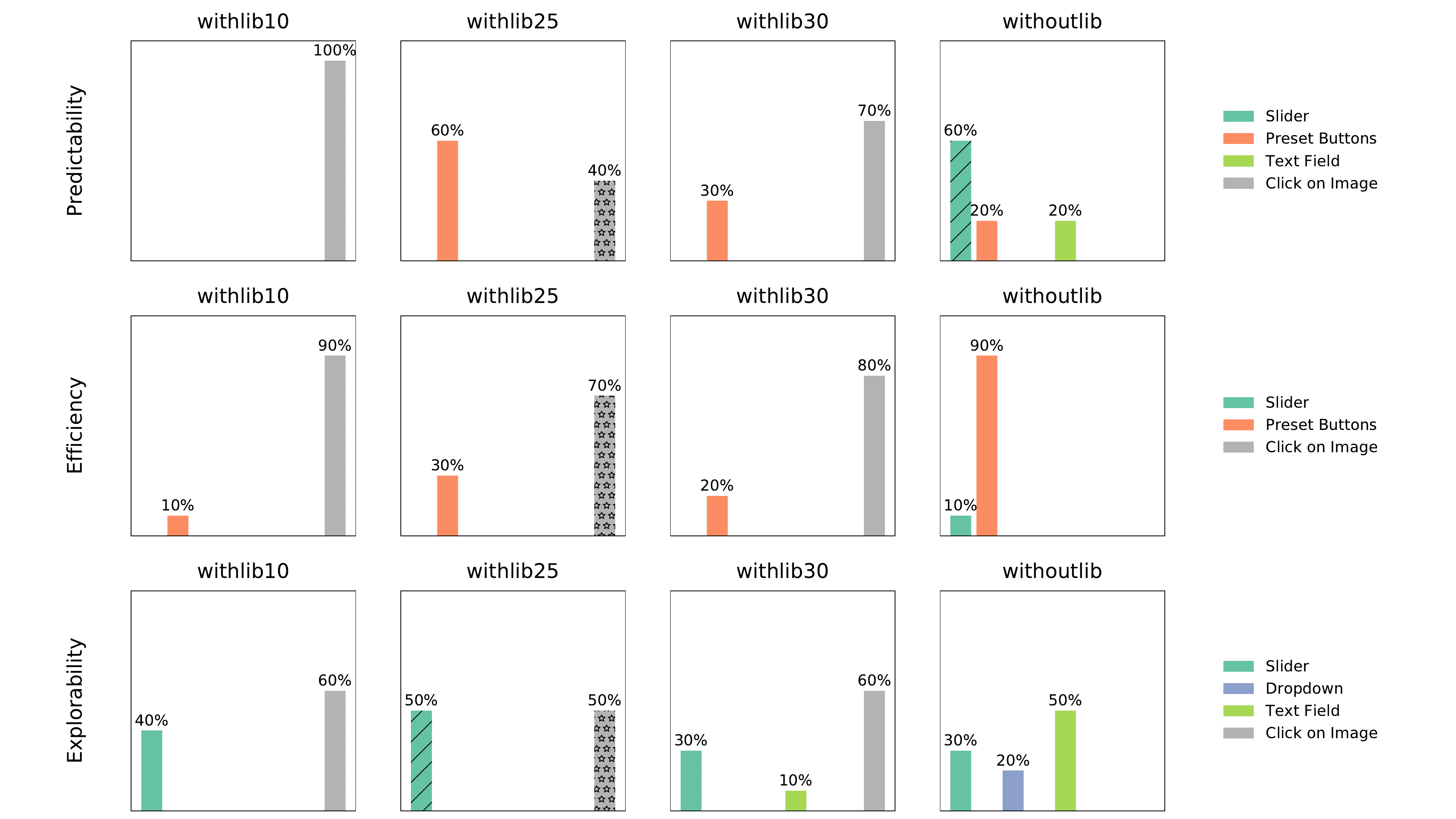}
    \caption{LLM-reasoned UI widgets for \texttt{design\_align\_text} with three sizes of the crowdsourced libraries and without any library. \texttt{Withlib10}, \texttt{withlib25}, and \texttt{withlib30} refer to using the crowdsourced libraries with 10, 25, and 30 user responses for all aspects and tasks, while \texttt{withoutlib} refers to using no library.}
    \label{fig:reasoning_design_align_text}
\end{figure*}

\begin{figure*}[!ht]
    \centering
    \includegraphics[width=0.85\textwidth]{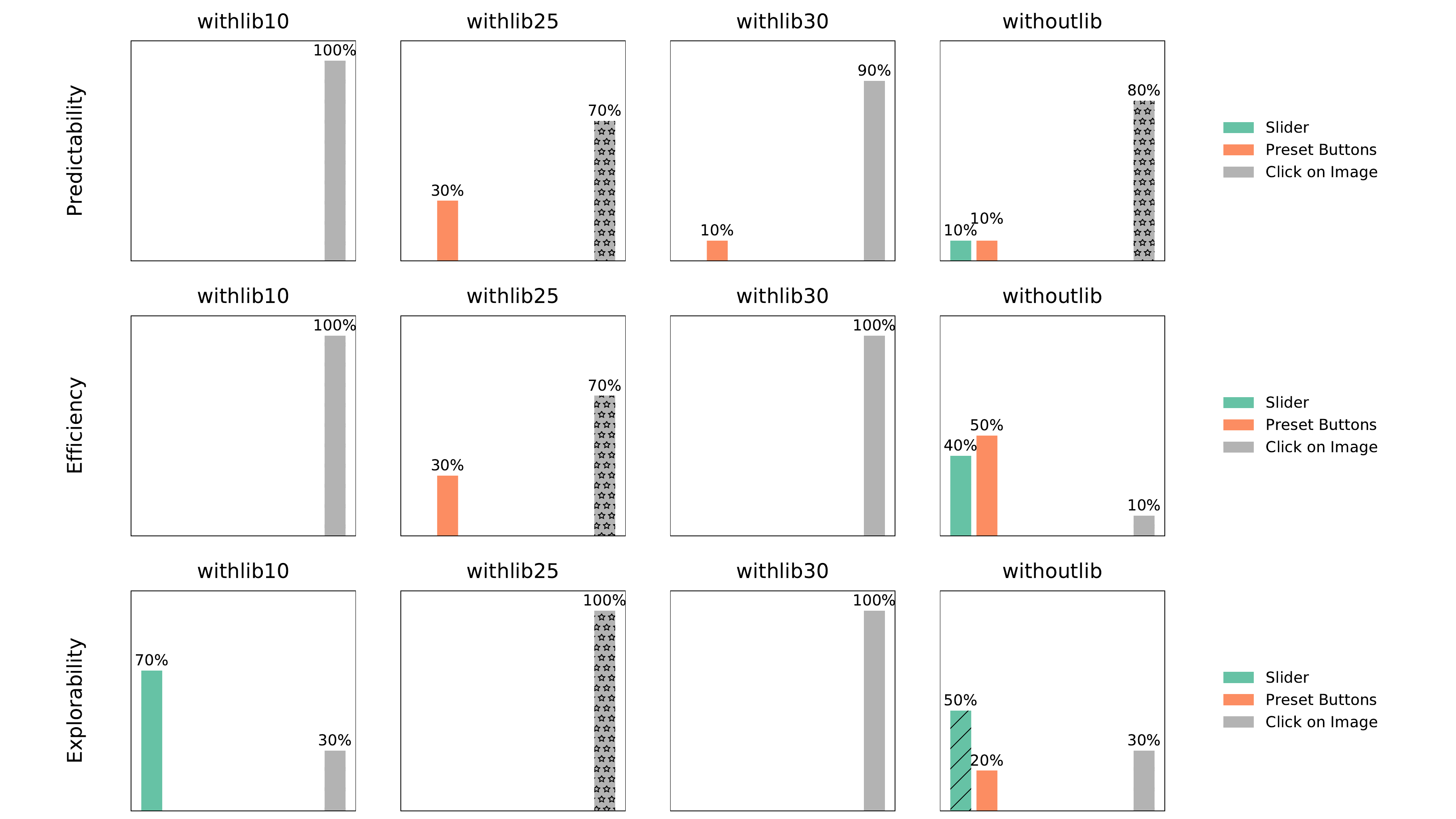}
    \caption{LLM-reasoned UI widgets for \texttt{design\_position\_logo} with three sizes of the crowdsourced libraries and without any library. \texttt{Withlib10}, \texttt{withlib25}, and \texttt{withlib30} refer to using the crowdsourced libraries with 10, 25, and 30 user responses for all aspects and tasks, while \texttt{withoutlib} refers to using no library.}
    \label{fig:reasoning_design_position_logo}
\end{figure*}

Figures \ref{fig:reasoning_image_adjust_exposure} to \ref{fig:reasoning_design_position_logo} show the LLM-reasoned UI widgets with crowdsourced libraries and without any library for the user evaluation tasks.

\clearpage

\section{Prompts for LLM UI Widget Reasoning and Code Generation}

\subsection{Prompts for UI Widget Reasoning with Crowdsourced UI Preference Library}
\label{appendix:prompt_widget_reasoning_withlib}
\begin{lstlisting}[language=json]
Based on the crowdsourced UI widget preference library, reason UI widget type for the user task. You should follow the steps below for the reasoning.

First, please take the definitions below:
    - Predictability: allows users to obtain results with no surprises or doesn't require users to deduce how to perform the interaction.
    - Efficiency: allows users to perform tasks with a minimum amount of effort.
    - Explorability: allows users to explore multiple possibilities and perform functions with high flexibility.

Second, information on the crowdsourced UI widget preference library is in the prompt:
    - Crowdsourced UI widget preference library task description: detailed descriptions of all the tasks.
    - Crowdsourced UI widget preference library widget frequency: the frequency of user-preferred widgets. Large numbers mean the corresponding widget is preferred by more people. 
    - Crowdsourced UI widget preference library widget reasons: the reasons for user-preferred widgets.

Third, search for the most relevant tasks from the crowdsourced UI widget preference library. 
    - You can compare the given task and the tasks names in the library, and refer to their task descriptions to help you to find the relevant tasks.
    - Additional task information:
        - "image_adjust_lightness", "image_adjust_saturation", and "image_adjust_hue" represent tasks that allow continuous value adjustment and discrete value selection, and "image_adjust_hue" is also related to color adjustment.
        - "image_adjust_fall_color", "image_color_match", and "image_adjust_color_balance" are related to color adjustment.
        - "image_place_watermark" and "image_place_vignette" are related to object positioning and discrete value selection.

Fourth, reason the most proper UI widget for predictability, efficiency, and explorability. 
    - Your reasoning should be based on the relevant tasks you found in the library.
    - After you find the relevant tasks, refer to the content of "Predictability", "Efficiency", or "Explorability" in the widget frequency and widget reasons.
    - You must refer to the widgets of high frequencies of the relevant tasks in widget frequency.
    - You must refer to widget reasons to help your reasoning.
    - The UI widget you reason must come from the given library.

Lastly, based on your reasoning, write down UI widgets for predictability, efficiency, and explorability in JSON format. 

Refer to the example below to provide your response. 
    - Replace the placeholders marked by <> with your response. Do not include <> in your response. 
    - You must keep all the existing information from the example and only replace placeholders.
    - The response must be in JSON format.
{
    "reasoning": {```reasoning
        "relevant tasks from the library": "<your reasoning>"

        "predictability_reasoning": {
            "<UI widget type>": "<your reasoning>"
        }
        
        "efficiency_reasoning": {
            "<UI widget type>": "<your reasoning>"
        }
        
        "explorability_reasoning": {
            "<UI widget type>": "<your reasoning>"
        }
    ```}
    
    "widget": {```widget
        "<task_name>": {
            "predictability": "<UI widget type>",
            "efficiency": "<UI widget type>",
            "explorability": "<UI widget type>"
        }
    ```}
}
\end{lstlisting}

\subsection{Prompts for UI Widget Code Generation}
\label{appendix:prompt_widget_coding}
\begin{lstlisting}[language=json]
Generate code for the UI widgets to perform the specified task. You should follow the steps below for the coding.

First, the UI widget code you provide must allow users to perform the task specified in the content of User task.

Second, the UI widget code you provide must follow the examples offered below for the implementation of Slider, Dropdown, Radio Buttons, Text Field, Preset Buttons, Color Wheel, Color Picker, Click on Image.
    - The UI widget types are in widget_type.
    - Find the example code for the relevant UI widget types.
    - When coding, you must only change the task to the specified task, implement the specified widgets, and keep the remaining code format the same as the example code.

Third, write your response in JSON format following the example responses below. 
{
    "task": "Adjust image hue",
    
    "widget_type": "Slider, Dropdown, Radio Buttons, Text Field, Preset Buttons, Color Wheel, Color Picker",

    "widget_code": {```python
\end{lstlisting}
\begin{lstlisting}[language=Python]
import numpy as np
import matplotlib.pyplot as plt
import ipywidgets as widgets
from IPython.display import display, clear_output
from PIL import Image
from skimage import data, img_as_ubyte
from matplotlib.patches import Wedge
import matplotlib.colors as mcolors

image = data.astronaut()
image = Image.fromarray(img_as_ubyte(image))

# Function to allow performing the task
def adjust_hue(image, hue):
    img_hsv = image.convert('HSV')  
    np_img = np.array(img_hsv)  

    hue_shift = int(hue * 255)
    

    np_img = np_img.astype(np.int32)
    np_img[..., 0] = (np_img[..., 0] + hue_shift) % 256

    np_img = np.clip(np_img, 0, 255).astype(np.uint8)

    adjusted_img = Image.fromarray(np_img, mode='HSV').convert('RGB') 
    return adjusted_img

# Function to create widgets
def create_hue_widgets():
    # Slider
    slider_label = widgets.Label(value='Slider:')
    slider = widgets.FloatSlider(value=0.0, min=0.0, max=1.0, step=0.01)

    # Dropdown
    dropdown_label = widgets.Label(value='Dropdown:')
    dropdown = widgets.Dropdown(options=[0.0, 0.2, 0.4, 0.6, 0.8], value=0.0)

    # Radio Buttons
    radio_buttons_label = widgets.Label(value='Radio Buttons:')
    radio_buttons = widgets.RadioButtons(options=[0.0, 0.2, 0.4, 0.6, 0.8], value=0.0)

    # Text Field
    text_field_label = widgets.Label(value='Text Field:')
    text_field = widgets.BoundedFloatText(value=0.0, min=0.0, max=1.0, step=0.01)
    
    # Preset Buttons
    preset_label = widgets.Label(value='Preset buttons:')
    
    preset_hues = [
        (0.0, 'red'), 
        (0.2, 'green'), 
        (0.4, 'cyan'), 
        (0.6, 'blue'), 
        (0.8, 'magenta')
    ]
    
    hue_mapping = {f"{hue}": hue for hue, color in preset_hues}
    
    preset_buttons = [
        widgets.Button(
            description=f"{hue}",
            layout=widgets.Layout(width='75px', height='30px'),
            style={'button_color': color}
        )
        for hue, color in preset_hues
    ]
    
    preset_buttons_box = widgets.HBox(preset_buttons)
    
    # Color Wheel
    color_wheel_label = widgets.Label(value='Color Wheel:')
    color_wheel = widgets.Output()
    create_color_wheel(color_wheel)
    
    # Color Picker
    color_picker_label = widgets.Label(value='Color Picker:')
    color_picker = widgets.ColorPicker(concise=True, value='#ffffff', disabled=False)

    # Layout
    layout = widgets.Layout(align_items='flex-start')
    spacer = widgets.Box(value='', layout=widgets.Layout(height='20px'))

    # Combine widgets into a vertical box
    widgets_box = widgets.VBox([slider_label, slider, spacer, 
                                dropdown_label, dropdown, spacer, 
                                radio_buttons_label, radio_buttons, spacer, 
                                text_field_label, text_field, spacer, 
                                preset_label, preset_buttons_box, spacer, 
                                color_wheel_label, color_wheel, spacer, 
                                color_picker_label, color_picker, spacer], layout=layout)
    
    return widgets_box, slider, dropdown, radio_buttons, text_field, preset_buttons, hue_mapping, color_wheel, color_picker

# Function to create and display the color wheel
def create_color_wheel(output):
    with output:
        clear_output(wait=True)
        
        fig, ax = plt.subplots(figsize=(1.5, 1.5))
        
        num_colors = 360
        theta = np.linspace(0, 2 * np.pi, num_colors, endpoint=False)
        colors = plt.cm.hsv(theta / (2 * np.pi))
        
        for i in range(num_colors):
            wedge = Wedge(center=(0, 0), r=1, theta1=(i * 360 / num_colors), 
                          theta2=((i + 1) * 360 / num_colors), color=colors[i], 
                          transform=ax.transData._b, clip_on=False)
            ax.add_patch(wedge)
        
        ax.set_aspect('equal')
        ax.set_xlim(-1.1, 1.1)
        ax.set_ylim(-1.1, 1.1)
        ax.axis('off')

        fig.canvas.mpl_connect('button_press_event', on_color_wheel_click)
        
        plt.show()

# Function to handle color wheel click
def on_color_wheel_click(event):
    if event.inaxes:
        x, y = event.xdata, event.ydata
        theta = np.arctan2(y, x) % (2 * np.pi)
        hue = theta / (2 * np.pi)
        hue = round(hue, 2) 

        update_plot(None, image, output, slider, dropdown, radio_buttons, text_field, preset_buttons, hue_mapping, hue=hue)

# Function to convert hex color to hue value
def hex_to_hue(hex_color):
    rgb = np.array([int(hex_color[i:i+2], 16) for i in (1, 3, 5)]) / 255.0
    hsv = mcolors.rgb_to_hsv(rgb.reshape(1, 1, 3))
    hue = hsv[0, 0, 0]
    return hue

# Function to update the image display based on widget values
def update_plot(change, image, output, slider, dropdown, radio_buttons, text_field, preset_buttons, hue_mapping, hue=None):
    with output:
        clear_output(wait=True)

        if hue is None:
            if change and change['owner'] in preset_buttons:
                clicked_button = change['owner']
                hue = hue_mapping[clicked_button.description]
            elif change and change['owner'] == slider:
                hue = slider.value
            elif change and change['owner'] == dropdown:
                hue = dropdown.value
            elif change and change['owner'] == radio_buttons:
                hue = radio_buttons.value
            elif change and change['owner'] == text_field:
                hue = text_field.value
            elif change and change['owner'] == color_picker:
                hue = hex_to_hue(change['new'])
            else:
                hue = 0.0

        adjusted_image = adjust_hue(image, hue)
        plt.figure(figsize=(4, 4))
        plt.imshow(adjusted_image)
        plt.axis('off')
        plt.title(f'Hue Adjustment: {hue:.2f}')
        plt.show()

# Function to link widgets to the update function
def link_widgets_to_update(image, output, slider, dropdown, radio_buttons, text_field, preset_buttons, hue_mapping, color_picker):
    def callback(change):
        update_plot(change, image, output, slider, dropdown, radio_buttons, text_field, preset_buttons, hue_mapping)
    
    slider.observe(callback, names='value')
    dropdown.observe(callback, names='value')
    radio_buttons.observe(callback, names='value')
    text_field.observe(callback, names='value')
    color_picker.observe(callback, names='value')
    for btn in preset_buttons:
        btn.on_click(lambda btn: update_plot({'owner': btn}, image, output, slider, dropdown, radio_buttons, text_field, preset_buttons, hue_mapping))

# Function to display the widgets and output plot
def display_hue_widgets():
    global output, slider, dropdown, radio_buttons, text_field, preset_buttons, hue_mapping, color_picker
    output = widgets.Output()
    
    widgets_box, slider, dropdown, radio_buttons, text_field, preset_buttons, hue_mapping, color_wheel, color_picker = create_hue_widgets()
    
    layout = widgets.Layout(width='100%', display='flex', justify_content='space-between')
    ui_and_image_box = widgets.HBox([widgets_box, output], layout=layout)
    
    display(ui_and_image_box)
    
    link_widgets_to_update(image, output, slider, dropdown, radio_buttons, text_field, preset_buttons, hue_mapping, color_picker)
    
    update_plot(None, image, output, slider, dropdown, radio_buttons, text_field, preset_buttons, hue_mapping)

display_hue_widgets()
\end{lstlisting}
\begin{lstlisting}[language=json]
    ```}
}
\end{lstlisting}


\section{User Evaluation Briefing and Instructions}
\label{appendix:user_eval_briefing_instructions}
\begin{lstlisting}[language=markdown]
Welcome!

This study evaluates user preferences for different UI widgets.

Your participation is voluntary. You are free to quit the study at any time. You can participate in this study only once.

All your responses (including demographic info) will be kept private and only used for research purposes.

# Study Instructions

- You will be working on 3 tasks. Each task is followed by 6 questions asking your UI widget preference by selecting one of two provided options for each question.

- For each task, please follow the task description and interact with **all the provided UI widgets**. 

- After interacting with the UI widgets, you will choose the preferences that make the most sense to you regarding the widgets' **predictability**, **efficiency**, or **explorability**.
    - Predictability: Allows you to obtain results with no surprises or doesn't require you to deduce how to perform the interaction.
    - Efficiency: Allows you to perform tasks with a minimum amount of effort.
    - Explorability: Allows you to explore multiple possibilities and perform functions with high flexibility.

- When making your selection, please consider the following **comprehensive factors**:
    - **Widget types**: Whether the widget types make sense to you regarding predictability, efficiency, or explorability.
    - **Scores**: How much each widget is recommended in terms of predictability, efficiency, or explorability.
        - Each widget type has a score out of 10.
        - In each option, a widget with higher scores indicates a stronger recommendation.
        - The total score of all widgets in each option adds up to 10.
    - **Reasons**: Why each widget is a good choice to allow predictability, efficiency, or explorability.
        - Some reasons may refer to users' UI widget preferences when performing other relevant tasks. These reasons are based on a library. Assume that such a library exists.
        - If a reason is based on the library, you will need to assess how well the information from this library is used for reasoning.

- **Do NOT** base your preference selection on:
    - The number of widget types, e.g., choosing an option simply because it offers more widget types.
    - The length of reasoning, e.g., choosing an option simply because the reasons are longer.

- Remember, your preference selection should consider widget types, scores, and reasons **collectively**. You can go back and interact with the UI widgets if you want.
    - If widget types are the same, consider the scores and reasons to make your selection.
    - If widget types and scores are the same, consider the reasons to make your selection.

\end{lstlisting}

\end{document}